\documentclass[oldversion]{aa}

\usepackage{times}
\usepackage{graphicx}
\usepackage{xspace}
\usepackage{epsfig}
\usepackage{natbib}
\usepackage{rotating}
\usepackage{dcolumn}

\newcommand{\futau}{FU\,Tau\,A\xspace}

\def\gsim{\;\lower4pt\hbox{${\buildrel\displaystyle >\over\sim}$}\,}
\def\lsim{\;\lower4pt\hbox{${\buildrel\displaystyle <\over\sim}$}\,}

\begin{document}

\title{X-Shooter spectroscopy of \futau \thanks{Based on observations collected at the Very Large Telescope of the European Southern Observatory under program 086.C-0173(A).}}

\author{B. Stelzer \inst{1} \and J.M. Alcal\'a \inst{2} \and A. Scholz \inst{3} \and A. Natta \inst{3,4} \and S. Randich \inst{4} \and E. Covino \inst{2}}

\offprints{B. Stelzer}

\institute{INAF - Osservatorio Astronomico di Palermo,
  Piazza del Parlamento 1,
  I-90134 Palermo, Italy \\ \email{B. Stelzer, stelzer@astropa.inaf.it} \and
  INAF - Osservatorio Astronomico di Capodimonte, 
  Via Moiariello 16, 
  I-80131 Napoli, Italy \and
  School of Cosmic Physics, 
  Dublin Institute for Advanced Studies,  
  31 Fitzwilliam Place, Dublin 2, Ireland \and 
  INAF - Osservatorio Astrofisico di Arcetri,
  Large E. Fermi 5, 
  I-50125 Firenze, Italy
}


\date{Received $<$date$>$ / Accepted $<$date$>$}

\abstract{We have analyzed a broad-band optical and near-infrared spectrum of \futau, 
a presumed young brown dwarf in the Taurus star forming region 
that has intrigued both theorists and observers by its overluminosity in the
HR diagram with respect to standard pre-main sequence evolutionary models. 
\futau is brighter than any other Taurus member of same or similar spectral type, 
and various phenomena (accretion, activity, binarity) have been put forth as a possible 
explanation.
The new data, obtained with the X-Shooter spectrograph at the Very Large Telescope, 
include an unprecedented wealth of information on stellar parameters and simultaneously
observed accretion and outflow indicators for \futau. 
We present the first measurements of gravity ($\log{g} = 3.5 \pm 0.5$),
radial velocity ($RV = 22.5 \pm 2.9$\,km/s), rotational velocity ($v \sin{i} = 20 \pm 5$\,km/s)
and lithium equivalent width ($W_{\rm Li} = 430 \pm 20$\,m\AA) for \futau. From the
rotational velocity and the published period we infer a disk inclination of $i \gsim 50^\circ$.
The lithium content is much lower than theoretically  
expected for such a young very low mass object, 
adding another puzzling feature to this object's properties. 
We determine the mass accretion rate of \futau from comparison of the luminosities of
$24$ emission lines to empirical calibrations from the literature and find
a mean of $\log{\dot{M}_{\rm acc}}\,{\rm [M_\odot/yr]} = -9.9$ with standard deviation 
$\sigma = 0.2$. 
The accretion rate determined independently from modeling of the excess emission
in the Balmer and Paschen continua is consistent with this value. The corresponding accretion
luminosity is too small to make a significant contribution to the bolometric luminosity.  
Strong magnetic activity affecting the stellar parameters or binarity of \futau, both 
combined with extreme youth, may be responsible for its position in the HR diagram. 
The existence of
an outflow in \futau is demonstrated through the first detection of forbidden emission lines
from which we obtain an estimate for the mass loss rate, 
$\log{\dot{M}_{\rm out}}\,{\rm [M_\odot/yr]} < -10.4$. 
The mass outflow and inflow rates can be combined to yield 
$\dot{M}_{\rm out}/\dot{M}_{\rm acc} \sim 0.3$, a value that is in agreement with jet launching models.

}
{
}
{
}
{
}
{
}
{
}


\maketitle

\section{Introduction}\label{sect:intro}

\futau was discovered by \cite{Luhman09.1} as the primary component of
a young brown dwarf (BD) binary in the Barnard\,215 cloud. This dark cloud is 
considered to be part of the Taurus molecular cloud complex, and a distance
of $140$\,pc is assigned \citep[e.g.][]{Torres09.0}. 
According to mid-infrared images from {\em Spitzer},
only one other young star, FT\,Tau, is present within $0.5^\circ$ of FU\,Tau
\citep{Luhman09.2}. 
Its isolated position makes FU\,Tau a benchmark object for studies of 
brown dwarf formation scenarios as most mechanisms, such as e.g. ejection
from a proto-stellar cluster, photo-evaporation, or 
disk fragmentation require the presence of higher-mass stars, 
see e.g. \cite{Whitworth07.0}.  

\cite{Luhman09.1} obtained optical and near-IR low-resolution spectroscopy
for the primary (\futau) and an optical spectrum for the secondary. They 
derived spectral types of M7.25 and M9.25 for the two components, respectively. 
By comparison to spectral templates, 
they extracted differing values for the extinction from the optical 
($A_{\rm V} \sim 2$\,mag) and the near-IR ($A_{\rm V} < 1$\,mag) spectra
of the primary, \futau, 
and adopted the higher value for the calculation of its bolometric luminosity. 
Masses of $0.05\,M_\odot$ and $0.015\,M_\odot$ were 
inferred comparing the position of the two objects in the HR diagram to the 
evolutionary pre-main sequence models of \cite{Baraffe98.1} and \cite{Chabrier00.1}. 
However, as noted by \cite{Luhman09.1}, in the HR diagram 
\futau is located well 
above the youngest ($1$\,Myr) isochrone of these models. 
The secondary, FU\,Tau\,B, also seems younger than $1$\,Myr albeit by 
a less amount, contesting the general notion of coevality for the component in 
binaries \citep{Luhman10.0}. 

In this article, we aim at investigating the nature of the primary, \futau. 
Various signatures of youth are apparent in the available observations of
this object. The presence of a circumstellar disk was established from an 
analysis of its spectral energy distribution (SED), where excess emission 
over a photosphere of the same spectral type 
is observed in all four {\em Spitzer}/IRAC bands, i.e. at $3.6\mu$m, $4.5\mu$m,
$5.8\mu$m and $8.0\mu$m \citep{Luhman09.1}.  
Similarly, an excess in the blue part of the SED indicates ongoing
accretion. This is bolstered by the high equivalent width of H$\alpha$
emission measured in Luhman et al.'s low-resolution spectra ($93$\,\AA), 
as well as the large width of the H$\alpha$ profile in a medium-resolution
spectrum from Gemini analysed by \cite{Stelzer10.0}. 
The full-width at $10$\,\% of the
peak height was measured to be $350$\,km/s, distinctly higher than the canonical
limit of $200$\,km/s considered to represent 
the borderline between magnetic activity
and accretion-dominated H$\alpha$ emission in BDs \citep{Jayawardhana03.2}. 
Applying the
calibration provided by \cite{Natta04.2} to the H$\alpha$ $10$\,\% width
yielded a mass accretion rate 
of $3.5 \cdot 10^{-10}\,{\rm M_\odot/yr}$. 
\cite{Stelzer10.0} caution that the H$\alpha$ line is only marginally
resolved in the Gemini spectrum. However, the mass accretion rate derived
from the line flux of the He\,I\,5876 line in the same spectrum,
using the calibration of \cite{Herczeg08.1} is about a factor two higher than
the value obtained from H$\alpha$, providing further evidence for strong
accretion.  

Recently, the FU\,Tau binary was the target of a {\em Chandra} X-ray 
observation with the
aim to study magnetic activity in two coeval BDs of slightly different 
effective temperature and/or mass. Unexpectedly, while the secondary was
not detected, the primary showed very
strong and soft X-ray emission, reminiscent of the T\,Tauri star
TW\,Hya where the bulk
of X-rays is produced in accretion shocks rather than the stellar corona
\citep[e.g.][]{Kastner02.1}. 
Considering the untypically low X-ray temperature and this analogy with TW\,Hya, 
\futau may be the first BD where 
X-ray emission from accretion shocks has been detected 
\citep{Stelzer10.0}. 

The current observational picture of FU\,Tau presents a number of 
ambiguities.
First, the over-luminosity in the HR diagram, especially for the primary, allows for
different interpretations as discussed by \cite{Stelzer10.0, Scholz12.0}. 
It could be due to extreme youth, to the primary being an unresolved binary, 
or to a strong contribution from accretion to the luminosity of \futau. Alternatively, 
the inhibiting influence of magnetic fields and/or rotation onto 
convection might for given luminosity lead to smaller effective temperature 
and, consequently, a wrong mass estimate \citep{Chabrier07.1}. 
Indeed, the modulations in photometric time series reveal the presence of both
hot spots, i.e. accretion, and cool spots, i.e. magnetic activity \citep{Scholz12.0}. 
Secondly, the velocities derived in the H$\alpha$ profile are smaller than the
infall speeds suggested by the observed X-ray temperature, leaving a doubt on the 
interpretation of the origin of the X-ray emission in accretion shocks.  

Aiming at a better understanding of this puzzling BD, we have obtained
broad-band spectroscopy from the UV to the near-IR and multi-color time-series
photometry for \futau. The motivation for collecting these data was to seek for a 
better understanding of its accretion and activity characteristics as possible cause for
the overluminosity of \futau in the HR diagram through the study of
spectral signatures. The new photometry is part of our study of \futau's long-term variability. 
The observations and data analysis are described in 
Sect.~\ref{sect:data_analysis}. Stellar properties are derived in Sect.~\ref{sect:results}.
In Sects.~\ref{sect:outflows} and~\ref{sect:accretion} we examine the outflow and accretion
characteristics of \futau, and in Sect.~\ref{sect:discussion} we discuss our results.

\section{Observations and data reduction}\label{sect:data_analysis}

\futau was observed on Jan 11, 2011 with the X-Shooter spectrograph at the 
VLT. The data were acquired within the INAF/GTO time \citep{Alcala11.0}.
With its three spectrograph arms, 
X-Shooter provides simultaneous wavelength coverage from $300-2480$\,nm.
Slit widths of $1.0^{\prime\prime}/0.9^{\prime\prime}/0.9^{\prime\prime}$
were used in the UVB/VIS/NIR arms, respectively,
yielding spectral resolutions of $5100/8800/5600$.  
The total exposure time in each of the three spectrograph arms was $1800$\,sec. 
We obtained a signal-to-noise ratio of $1-8$ in the UVB, $10-20$ in the VIS, and 
$20-30$ in the NIR arm. 
The data were obtained in {\em nod} mode and were reduced independently for 
 each arm with the X-Shooter pipeline, 
v1.3.7 \citep{Modigliani10.0}. 
Following the standard steps including bias or dark subtraction, flat fielding,
optimal extraction, wavelength calibration, sky subtraction, correction for atmospheric
extinction and flux calibration. 
However, the pipeline flux-calibrated spectra are not corrected for telluric absorption 
bands. 
The telluric correction was performed independently in the VIS and NIR spectra
in the following way. 
For the VIS arm the spectrum of the telluric standard (HIP\,20789, spectral type B7\,V), 
normalized to the continuum, was used as input in the 
IRAF\footnote{IRAF is distributed by the National Optical Astronomy Observatories, which are operated by the Association of Universities for Research in Astronomy, Inc., under cooperative agreement with the National Science Foundation.} 
task "telluric". For the NIR spectrum, a response function was first 
derived by dividing the non flux-calibrated spectrum of the telluric 
by a black-body of the same effective temperature (T$_{\rm eff}$=13,000\,K)
as the telluric standard. This response function, containing the telluric 
lines, was then used as input in the IRAF task "telluric". In this way, 
the telluric correction and the correction for the response function 
were done simultaneously. Although the shape of the resulting spectrum after 
this procedure is correct, the flux calibration is only relative to the response function. 
To bring the NIR spectrum into an absolute flux scale we multiplied with a factor 
that was estimated using the pipeline flux-calibrated spectrum. 
Finally, wavelength shifts due to instrumental flexures were 
corrected using the {\em flexcomp} package within the 
pipeline. The precision of the wavelength calibration is 
better than 0.01\,pix corresponding to $0.002$\,nm in the UVB and VIS arms and to  
$0.006$\,nm in the NIR arm. 

\begin{figure*}
\begin{center}
\includegraphics[width=18cm]{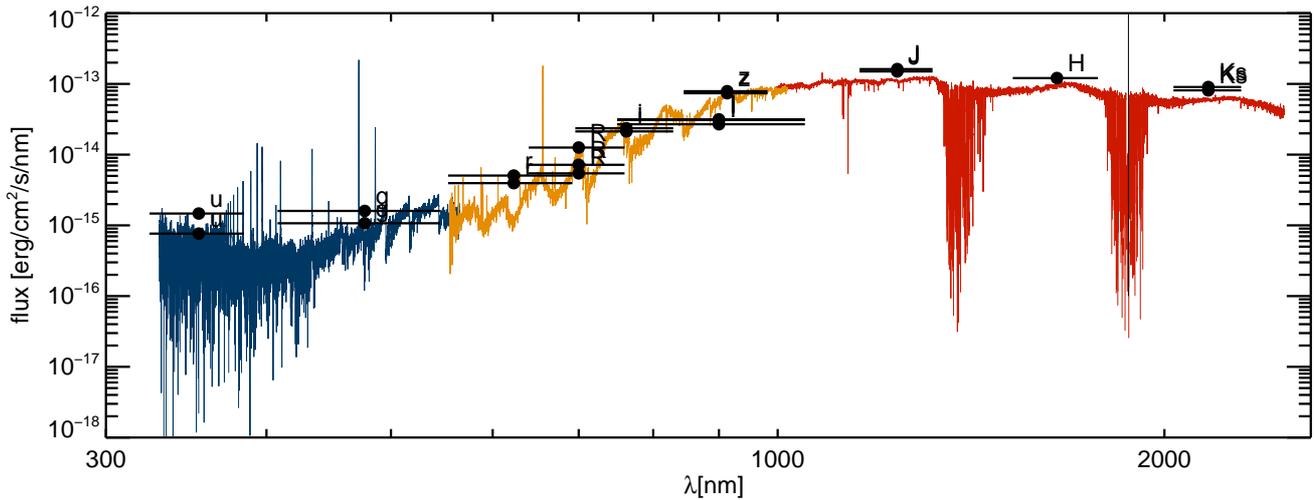}
\caption{Broad-band flux-calibrated X-Shooter spectrum of \futau.  
Published photometry from \protect\cite{Luhman09.1}, from 
\protect\cite{Scholz12.0} and unpublished $RIJK$ photometry obtained in September 2012 
with Andicam at the 1.3m telescope on Cerro Tololo as part of the 
SMARTS program DUBLIN-12A-001 is overplotted.}
\label{fig:spectrum_allarms}
\end{center}
\end{figure*}

The flux-calibrated spectrum of \futau for the full X-Shooter wavelength range 
is shown in Fig.~\ref{fig:spectrum_allarms}. 
Photometry from the literature is overlaid, demonstrating 
the high quality of the flux calibration. 
In particular, the match of the different instrumental arms is very good. In fact,
during the observation the seeing was excellent ($\approx 0.85^{\prime\prime}$). 

\section{Stellar properties}\label{sect:results}

\subsection{Spectral type and extinction}\label{subsect:results_spt}

\begin{figure}
\begin{center}
\includegraphics[width=9cm]{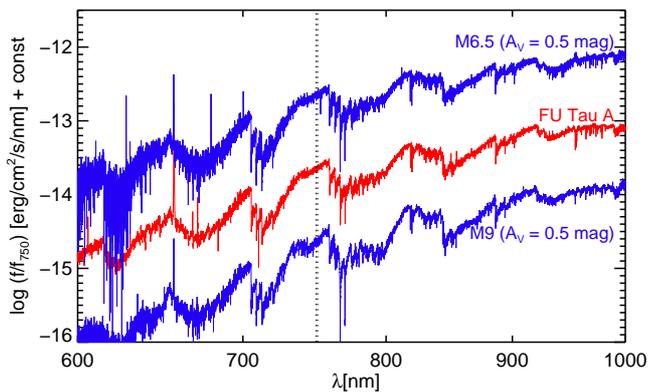}
\caption{X-Shooter VIS spectrum of \futau compared to two spectral templates from 
Manara et al. (2013). A reddening of $A_{\rm V} = 0.5$\,mag has been applied to the
templates. The spectra are normalized at $750$\,nm and the templates are
vertically offset.} 
\label{fig:spt}
\end{center}
\end{figure}

We estimated the spectral type and extinction 
of \futau by comparing it to the spectral templates 
defined by Manara et al. (2013, A\&A subm.) on the basis of X-Shooter spectra for $24$
non-accreting and unabsorbed YSOs in various star forming regions. These spectra define a 
continuous spectral sequence from M0 to M6.5 with steps of 0.5 in spectral subclass.
In addition they comprise two objects at the end of the M subclass. At their young ages 
($1-10$\,Myr) these objects are expected to have similar gravity to \futau. 
We determined simultaneously the spectral type and the extinction of \futau by 
artificially reddening the templates between $A_{\rm V} = 0...2$\,mag until the best match to 
\futau was found. 
For the reddening we used the extinction law of \cite{Weingartner01.0}. 
In Fig.~\ref{fig:spt}, the spectrum of \futau is compared to that of Par-Lup3-1 (M6.5) and
DENIS-P\,J124514.1-442907 (M9).
The spectrum of Par-Lup3-1 appears very similar to \futau and we conclude that 
the spectral type of \futau is M6.5 or slightly later. The gap in the spectral type
sequence of Manara et al. (2013) does not allow us to put stronger constraints.
We estimated an extinction of $A_{\rm V} \sim 0.5 \pm 0.5$\,mag. 
The templates in Fig.~\ref{fig:spt} are both reddened by this amount. 

In a second approach to determine the spectral type of \futau 
we have calculated various spectral indices following \cite{Riddick07.0} for
optical wavelengths, and the ${\rm H_2 O}$ index defined by \cite{Allers07.1} and the
${\rm H_2 O}$-K2 index from \cite{RojasAyala12.0} for the near-IR spectrum.
The indices that we have used are consistent with the spectral sequence 
of Manara et al. (2013). 
The flux ratios for the spectral indices were calculated from the spectrum of \futau 
after modifying it by various amounts of extinction between $A_{\rm V} = 0$ and $2$\,mag.
We confirm that the optical spectral indices from \cite{Riddick07.0} are independent on
extinction for the range of extinctions tested here  
and derive a spectral type of ${\rm M}6.6 \pm 0.4$ for FU\,Tau building the average 
and standard deviation of the VO2, R1, R2, R3, TiO8465, and VO7445 indices.
The ${\rm H_2 O}$ index yields 
M6.7 $\pm$ 0.2 
and the ${\rm H_2 O}$-K2 index 
M7.1 
for a range of $A_{\rm V} = 0...1$\,mag.  
\cite{Riddick07.0} and \cite{Allers07.1} provide calibrations specifically for 
{\em young} M dwarfs, i.e. their spectral indices are independent on gravity, and we find
excellent correspondence in the derived spectral types. The slightly later spectral type
obtained from the ${\rm H_2 O}$-K2 index may be related to the fact that \cite{RojasAyala12.0} 
have calibrated it for nearby (evolved) M dwarfs. 
 
The spectral type we obtained from the X-Shooter spectrum with the two methods 
explained above is 
also consistent
with our earlier results from low-resolution spectroscopy where we found values between 
M6.6 and M6.8 in a series of five spectra \citep{Scholz12.0}. Using a different set
of spectral templates \cite{Luhman09.1} found a similar, 
slightly cooler spectral type of M$7.25 \pm 0.25$. 
 
%


Luhman et al.'s value for the extinction derived from their optical spectrum 
($A_{\rm V} < 1$\,mag) is also similar to ours but they found 
a higher value from the near-IR spectrum ($A_{\rm V} = 2$\,mag). 
The latter value is clearly incompatible with the X-Shooter NIR spectrum
for which we derive $A_{\rm V} \sim 0.75 \pm 0.5$\,mag in an analogous way as 
the one described for the VIS. 
Throughout this paper we adopt 
the spectral type and extinction derived from the VIS spectrum.
We expect the determination of extinction using zero-extinction 
spectral templates observed with the same instrument to be preciser in the VIS
than in the NIR because, generally, the effects of extinction are larger in the VIS
than in the NIR.

\subsection{Stellar parameters}\label{subsect:results_params}

\begin{figure}
\begin{center}
\includegraphics[width=6cm,angle=270]{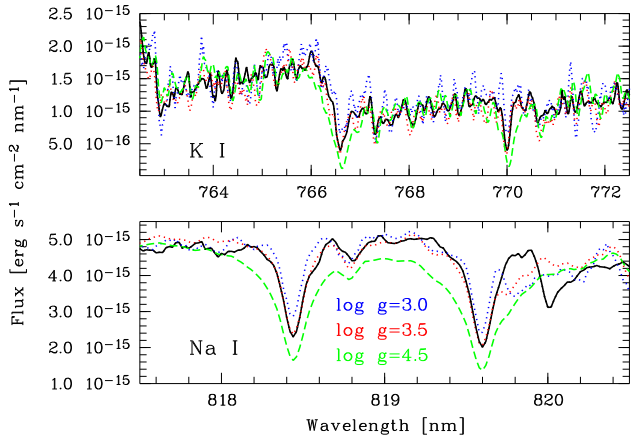}
\caption{Detail of the spectrum of FU\,Tau (black solid lines) 
       in the wavelength range around the \ion{K}{i} (upper panel) and
       \ion{Na}{i} (lower panel) absorption doublets. The spectrum of
       FU\,Tau has been corrected for telluric absorption lines.
       The red dotted line and the green dashed line represent synthetic spectra 
       with $\log{g}$ as labelled for $T_{\rm eff} = 2900$\,K and rotationally broadened
       to $v \sin{i} = 20$\,km/s.}
\label{fig:gravity}
\end{center}
\end{figure}

Our new measures for the spectral type and the extinction can be used
to revise the stellar parameters of \futau. 
The radius of \futau can be obtained from Stefan-Boltzmann's law. 
Rather than using the literature value for the bolometric luminosity we recompute it 
taking account of the difference in our adopted $A_{\rm V}$ and the value assumed by \cite{Luhman09.1}.
From the $J$ magnitude given by \cite{Luhman09.1} and a bolometric correction of 
$B.C._{\rm J} = 2.05$ \citep{Dahn02.1} we find $L_{\rm bol} = 0.13\,L_\odot$. 
We have measured a slightly earlier spectral type than \cite{Luhman09.1}, 
yielding an effective temperature 
of $T_{\rm eff} = 2940$\,K for a temperature scale intermediate between dwarfs and giants 
\citep{Luhman99.1}. For these numbers we derive $R_* = 1.4\,R_\odot$, substantially lower
than the value resulting from the luminosity and temperature given by \cite{Luhman09.1},
$R_* = 1.8\,R_\odot$. Extrapolating the position of \futau in the HR diagram down to 
the youngest isochrone of the evolutionary models by \cite{Chabrier00.1}, the mass is
$0.08\,M_\odot$, placing \futau right at the hydrogen burning mass limit. 

We used the gravity- and temperature-sensitive absorption doublets of 
\ion{Na}{i} at $\lambda\lambda$ 818.33, 819.48\,nm
and of \ion{K}{i} at $\lambda\lambda$ 766.48, 769.89\,nm to determine $\log{g}$ by
comparison of the X-Shooter data to synthetic spectra.
We retrieved BT-DUSTY 
model spectra for a range of $T_{\rm eff}$ around the expected value of
$2940$\,K and a range of $\log{g}$ values 
from the {\em star, brown dwarf \& planet atmosphere web simulator}
\citep[][]{Allard10.0}, electronically 
available\footnote{http://phoenix.ens-lyon.fr/simulator/index.faces}.
In Fig.~\ref{fig:gravity} the synthetic spectra for $T_{\rm eff} = 2900$\,K 
with three different values for $\log{g}$ are overlaid on the spectrum of 
\futau in the region of the Na and K doublets.
The model spectra have been rotationally broadened to $v \sin{i} = 20$\,km/s,
the rotation rate determined for \futau in Sect.~\ref{subsect:results_vsini}.
The width of the observed lines in Fig.~\ref{fig:gravity} is in good agreement with a gravity of
$\log{g} = 3.5 \pm 0.5$. The uncertainty is due to the $0.5$ steps for $\log{g}$
in the grid of synthetic spectra. The gravity derived from the spectrum 
agrees with the value expected from the evolutionary models
of \cite{Baraffe98.1} adopting the stellar parameters derived above and an age 
of $1$\,Myr.

\subsection{Lithium absorption}\label{subsect:results_lithium}

Theoretical models predict that the lithium content of low-mass stars 
gets rapidly depleted throughout the first $\sim 50$\,Myrs of the 
pre-main sequence evolution \citep[e.g.][]{dAntona97.0, Baraffe98.1}.
In particular, the Li\,I$\lambda$670.8\,nm absorption 
line is a well-known age indicator in young low-mass objects. This was 
shown, e.g., in studies of the line equivalent width for clusters and associations 
with different HR diagram ages \citep[e.g.][]{Mentuch08.0}. 
The age at which lithium depletes increases with decreasing mass. 
For the fully convective very low-mass ($< 0.35\,M_\odot$) 
objects in star forming regions standard evolutionary models predict that 
the original abundances are still retained until an age of at least $10$\,Myrs
\citep[e.g.][]{Jeffries06.0}  
and a cosmic abundance of A(Li) $\approx 3.1$ \citep{ZapateroOsorio02.2} is expected. 

\begin{figure}
\begin{center}
\includegraphics[width=6cm, angle=270]{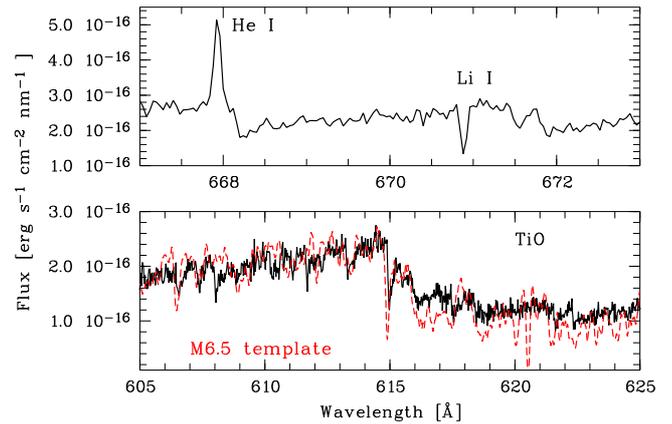}
\caption{\ion{Li}{i}$\lambda 670.8$\,nm 
absorption in \futau (top) and TiO absorption band at $615$\,nm, 
a diagnostic of veiling, compared to the same spectral region in the unveiled 
Class\,III template Par-Lup3-1 (bottom).}
\label{fig:veiling}
\end{center}
\end{figure}

The Li\,I $\lambda 670.8$\,nm absorption of \futau is shown in the top panel of 
Fig.~\ref{fig:veiling}. We determined an equivalent width of $0.43 \pm 0.02$\,\AA. 
In practice, for accreting objects photospheric absorption lines may be filled in by excess 
continuum emission leading to shallower lines and an underestimate of the line equivalent widths. 
The amount of this `veiling' can be estimated by comparison with the spectrum of an
unveiled star of the same spectral type. 
The only available template for \futau observed with the same instrument and the same 
spectral resolution is Par-Lup3-1 (M6.5). 
We have thus measured the EWs of the TiO absorption bands in the region between $660$\,nm 
and $680$\,nm \citep{Valenti98.1} for both the template and \futau.  
We calculated the ratio $EW_{\rm template}/EW_{\rm FUTauA}$ 
for eight different bands and found 
that the ratio is at most $1.2$. A more detailed evaluation is prohibited by the
noise level of the template spectrum. If these differences
in EWs are due to veiling the corrected EW(Li) value for 
\futau would be $\lsim 516$\,m\AA. 
We have repeated the same exercise using the spectral
features shown in the bottom panel of Fig.~\ref{fig:veiling}. The result
is a correction factor of $\leq 1.3$ which is consistent
with weak veiling.

%
%

\begin{figure}
\begin{center}
\includegraphics[width=9cm]{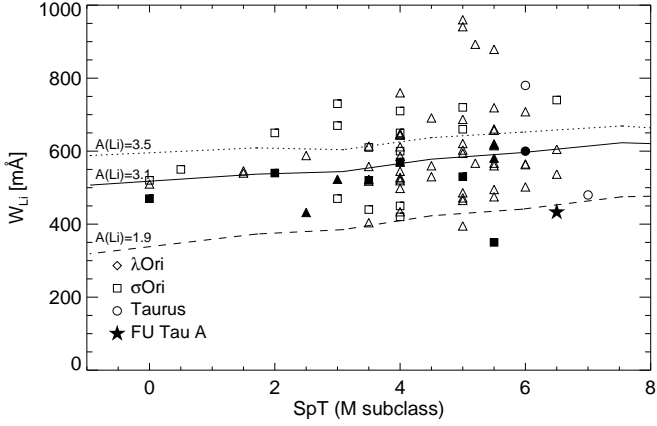}
\caption{Lithium equivalent width versus spectral type for \futau (filled star symbol)
and other measurements of M-type YSOs collected from the literature (see text in 
Sect.\ref{subsect:results_lithium}). Accretors are characterized by filled symbols.
Also plotted are the curves of growth for $\log{g} = 4.0$ and three different abundances with labels, 
calculated by \cite{ZapateroOsorio02.2}.}
\label{fig:lithium}
\end{center}
\end{figure}

In Fig.~\ref{fig:lithium} we compare our measurement to other studies of lithium in YSOs
of late-M spectral type from the literature: 
$\sigma$\,Ori \citep{ZapateroOsorio02.2}, $\lambda$\,Ori \citep{Bayo11.0},
and three very low-mass (VLM) objects in Taurus \citep{Barrado04.3}. 
A compilation of lithium equivalent width measurements in Taurus is given by \cite{Sestito08.0}
for a larger range of spectral types. We do not consider this sample here because it 
includes a number of stars with dubious membership.  
For the case of $\lambda$\,Ori we consider only the subsample with confirmed cluster membership
and with spectroscopically determined spectral types. 

Accreting objects are highlighted with filled plotting symbols in Fig.~\ref{fig:lithium}
and have been selected as follows: In $\sigma$\,Ori we consider accreting all objects with 
$W_{\rm H\alpha} > 20$\,\AA, in $\lambda$\,Ori we rely on the `accretor flag' given by
\cite{Bayo11.0} which is based on H$\alpha$ emission and in Taurus we consider accreting
objects with $W_{\rm H\alpha 10\%} > 300$\,km/s. The choice of the $10$\,\% width rather than
the equivalent width as accretion diagnostic for Taurus is motivated by the fact that one
object, KPNO-Tau\,5, has $W_{\rm H\alpha} = 21.1$\,\AA, at the borderline to the accretion
regime but is classified as non-accretor due to the narrow H$\alpha$ profile and
weak He\,I $\lambda$ 667.8 emission.

The above-mentioned star forming regions all have an age
of $1-5$\,Myr and the lithium content of M stars is expected to reflect the initial cosmic
abundance, A(Li) = 3.1. However, a large spread of equivalent widths is observed for a 
given spectral type. 
Observations obtained with different spectral resolution may result in systematic differences
of the line measurements but these errors are smaller than the observed spread of the 
equivalent width values.
This spread also corresponds to an unexplained spread in abundances as evident by the
over-plotted curves of growth for $\log{g} = 4.0$ calculated by \cite{ZapateroOsorio02.2}.
\futau presents the lowest Li measurement observed so far for its spectral class.
Taking into account veiling, \futau moves up in the diagram but is still well below the
theoretically expected abundance. 

\subsection{Radial velocity and UVW velocity}\label{subsect:results_rv}

We have used the strongest absorption lines, 
the \ion{K}{I} doublet at $\lambda\lambda$ 766.48, 769.89\,nm
and the \ion{Na}{i} doublet at $\lambda\lambda$ 818.33, 819.48\,nm,  
to estimate the radial velocity of \futau by comparison of the observed line centers
to the rest wavelengths extracted from the 
NIST Atomic Spectra Database\footnote{http://www.nist.gov/pml/data/asd.cfm/}. 
After application of the barycentric correction ($-20.6$\,km/s)
the mean Doppler shift of these lines yields 
$RV = 22.5 \pm 2.9$\,km/s  
where we have added to the standard deviation of the four measured lines the
uncertainty of the wavelength calibration (see Sect.~\ref{sect:data_analysis}).  
The local-standard-of-rest velocity of \futau is $12.5 \pm 2.9$\,km/s. 

\cite{Bertout06.1} have compiled a list of $127$ stars in the Taurus star forming
complex with $RV$ measurements. These authors note that 
no systematic high-precision $RV$ survey has been performed in Taurus. 
Their $RV$ distribution is strongly peaked at $15-16$\,km/s,
in agreement with historic smaller samples presented by \cite{Hartmann86.1} and \cite{Walter88.1}.
About $\sim 5$\,\% of this list have $RV \geq 22$\,km/s, i.e. the $RV$ of \futau is
only marginally compatible with that of Taurus. 

We have combined our $RV$ measurement with the proper motion of \futau given in the 
literature to obtain the $UVW$ velocities. We use both the proper motion values of \cite{Ducourant05.0}
and those of \cite{Luhman09.1}. The resulting galactic velocity components for a distance of 
$140$\,pc are listed in Table~\ref{tab:uvw}.
The average space motion for Taurus for the same distance is 
$(U,V,W) = (-16.5 \pm 4.6,-13.2 \pm 2.5,-11.0 \pm 4.0)$\,km/s \citep{Bertout06.1}.
\futau is in rough agreement with these values. 

\begin{table}
\begin{center}
\caption{Proper motion of \futau from the literature and galactic velocity calculated 
using our $RV$ measurement and $d = 140$\,pc.}
\label{tab:uvw}
\begin{tabular}{lrr}\hline
Reference & \multicolumn{1}{c}{$\mu_{\rm \alpha}$,$\mu_{\rm \delta}$} & \multicolumn{1}{c}{$(U,V,W)$} \\ 
          & \multicolumn{1}{c}{[mas/yr]}            & \multicolumn{1}{c}{[km/s]} \\ \hline
\cite{Ducourant05.0} & $14.7$,$-26.0$ & ($-22.09$,$-9.08$,$-10.36$) \\
\cite{Luhman09.1}    & $ 7.2$,$-17.5$ & ($-23.54$,$-16.55$,$-10.41$) \\
\hline
\end{tabular}
\end{center}
\end{table}

\subsection{Rotational velocity}\label{subsect:results_vsini}

\begin{figure}
\begin{center}
\includegraphics[width=6cm,angle=270]{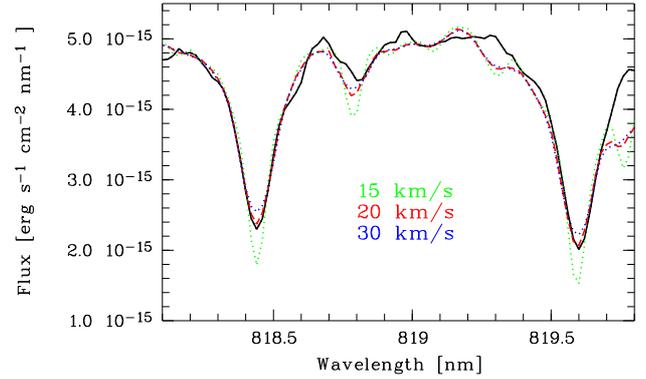}
\caption{The \ion{Na}{i} absorption doublet $\lambda\lambda$\,818.33,819.48\,nm
of \futau is shown as continuous black line.
The synthetic spectrum, broadened at projected rotational
velocities of $15$, $20$ and $30$\,km/s, is represented with the
green-dotted, red-dashed and blue-dotted lines, respectively.}
\label{fig:vsini}
\end{center}
\end{figure}

The projected rotational velocity, $v \sin{i}$, was estimated by 
comparing the profile of the \ion{Na}{i} absorption doublet at 
$\lambda \lambda$818.33,819.48\,nm with that of a synthetic BT-DUSTY spectrum 
of the same effective temperature and gravity as \futau. The synthetic 
spectrum was gathered from the 
{\em star, brown dwarf \& planet atmosphere web simulator} 
\citep{Allard10.0} using a $v \sin{i}=0$\,km/s. In order 
to reproduce the profile of the \ion{Na}{i} absorption doublet in \futau, 
the synthetic spectrum was convolved with rotational profiles
\cite[see][]{Gray92.0} of several values of $v \sin{i}$.  The best 
match is for $v \sin{i}=20$\,km/s (see Fig.~\ref{fig:vsini}). We estimate an 
error on the order $5$\,km/s. 

We can combine our measurements of the rotational velocity ($v \sin{i}$) with 
the photometrically determined rotation period from \cite{Scholz12.0} ($P = 4.0 \pm 0.2$\,d)
and find $R \sin{i} = 1.6 \pm 0.5\,{\rm R_\odot}$. 
The value obtained for $R \sin{i}$ is compatible with our new value for the stellar radius
($R_* = 1.4\,{\rm R_\odot}$) within the errors and 
implies a disk inclination angle $i \geq 53^\circ$.

\section{Outflows}\label{sect:outflows}

We have searched for forbidden emission lines (FELs) as signatures of shocks forming in outflows.
A list of the equivalent widths (EWs) and line fluxes for the 
FELs detected in \futau is given in Cols.~3 and~4 of Table~\ref{tab:forbidden}. 
The EWs and fluxes are obtained from an by-eye estimate of the local continuum. 
The uncertainties of the EWs and the fluxes 
represent the mean and standard deviation from three measurements carried out on 
each line. 
In our approach of estimating these uncertainties we take account of the fact that the 
uncertainties in the flux are dominated by the uncertainty of the extinction while the 
major source of the uncertainties in the EWs are the statistical fluctuations.
Therefore, for the errors of the EWs we have measured the line in the spectrum 
dereddened by $0.5$\,mag at another two positions corresponding to our estimate of
the upper and lower value of the adjacent continuum emission and computed the 
standard deviation of the three measurements. 
For the uncertainties of the line fluxes we have measured the line in the 
observed spectrum (corresponding to an assumption of $A_{\rm V} = 0$\,mag) and 
in the spectrum dereddened for $A_{\rm V} = 1$\,mag, and combine those two values
with that from the spectrum dereddened by $0.5$\,mag to get the standard deviation.
\begin{table*} \begin{center}
\caption{Forbidden emission lines: nominal wavelength (col.2), line equivalent width and flux (cols.3 and~4), radial velocity with respect to the barycenter-corrected stellar velocity (col.~5), tangential velocity (col.~6) and uncertainty of the velocities.}
\label{tab:forbidden} \newcolumntype{d}[1]{D{.}{.}{#1}}
\begin{tabular}{lcccccc} \\ \hline
Element  &  $\lambda_0$ & $EW$ & $f_{\rm line} \cdot 10^{16}$ & $RV$ & $V_{\rm T}$ & $\Delta V$ \\
 & [nm] & [nm] & [${\rm erg/cm^2/s}$] & [km/s] & [km/s] & [km/s] \\ \hline
   SII & $ 406.860$ & $  -0.305 \pm   0.102$ & $   2.7 \pm   1.5$ & $ -26.6$ & $  99.2$ & $   3.2$ \\
   SII & $ 407.635$ & $  -0.081 \pm   0.083$ & $   0.4 \pm   0.2$ & ... & ... & ... \\
    OI & $ 630.030$ & $  -0.323 \pm   0.053$ & $   8.8 \pm   3.4$ & $ -12.6$ & $  47.2$ & $   3.2$ \\
    OI & $ 636.378$ & $  -0.067 \pm   0.011$ & $   2.7 \pm   1.0$ & ... & ... & ... \\
   NII & $ 658.260$ & $  -0.009 \pm   0.007$ & $   0.7 \pm   0.4$ & ... & ... & ... \\
   SII & $ 673.082$ & $  -0.030 \pm   0.010$ & $   1.2 \pm   0.4$ & $ -11.0$ & $  41.1$ & $   3.2$ \\
\hline \end{tabular} \end{center} \end{table*}

The non-detection of the [SII]$\lambda 671.6$\,nm line does not allow us to apply the
technique of \cite{Bacciotti99.0} for estimating plasma parameters from
the line ratios of FELs. 
For a detailed analysis of luminosities, kinematics and mass outflow rate ($\dot{M}_{\rm out}$)
of \futau we consider only the two most prominent lines, [OI]$\lambda$630.0\,nm and 
[SII]$\lambda$673.1\,nm, 
and we follow the prescription described by \cite{Hartigan95.1} who 
presented relations of the type
\begin{equation}
\log{\dot{M}_{\rm out}} = X \cdot (1 + \frac{n_c}{n_e})(\frac{V_{\rm T}}{150\,{\rm km s^{-1}}})(\frac{l_{\rm T}}{2 \cdot 10^{15}\,{\rm cm}})^{-1}(\frac{L_{\rm line}}{L_{\rm sun}})
\label{eq:mout}
\end{equation}
The numerical constant $X$ is derived in \cite{Hartigan95.1} 
for the [SII]$\lambda 673.1$\,nm and the [OI]$\lambda 630.0$\,nm lines.
Eq.~\ref{eq:mout} involves the electron density ($n_{\rm e}$),
the tangential velocity of the outflow ($V_{\rm T}$) and the projected size of the
aperture in the plane of the sky ($l_{\rm T}$). For a $1^{\prime\prime}$ slit we obtain
for the distance of Taurus $l_{\rm T} = 2 \cdot 10^{15}$\,cm.
For the reasonable assumption that the outflow is perpendicular to the disk we  
use the minimum possible disk inclination angle ($i_{\rm disk} \sim 50^\circ$; see 
Sect.~\ref{subsect:results_vsini}) and the measured $RV$ of the FELs with respect to the
stellar $RV$ 
to estimate the tangential velocities. The derived quantities are given in Table~\ref{tab:forbidden}. 
The uncertainty in the velocities are given in col.7. 
They include the standard deviation of three wavelength 
measurements and the uncertainty of the stellar $RV$. 

We have no measurement of the electron density for the outflow of \futau. 
A lower limit to $n_{\rm e}$ is given by the fact that we detect [SII]\,$\lambda$673.1\,nm
but not [SII]$\lambda$617.7\,nm emission. The flux ratio between these two lines is,
therefore, larger than unity and $n_{\rm e} > 10^3\,{\rm cm^{-3}}$ 
\citep[e.g.][]{Osterbrock89.0}. 
This gives an upper limit to the mass loss rate determined from 
[SII]$\lambda$673.1\,nm of $\log{\dot{M}_{\rm out}}\,{\rm [M_\odot/yr]} < -10.4$.
Assuming the same lower limit for the electron density, the [OI]$\lambda$630.0\,nm line 
gives a much less stringent constraint on $\dot{M}_{\rm out}$. 
We note that the mass loss rates computed from the two lines agree for 
$n_{\rm e} \sim 10^5\,{\rm cm^{-3}}$. For that case, we find 
$\log{\dot{M}_{\rm out}}\,{\rm [M_\odot/yr]} < -11.5$.
Recall also that the value of the inclination angle assumed in this calculation is the
minimum possible value on the basis of the rotational properties of \futau. If the disk
inclination is higher than that value, the upper limit on the mass loss rate is higher
and less meaningful, e.g. for $i = 75^\circ$ the mass loss rate obtained from [SII] is 
$<10^{-10.0}\,{\rm M_\odot/yr}$ for $n_{\rm e} > 10^3\,{\rm cm^{-3}}$ and 
$<10^{-11.0}\,{\rm M_\odot/yr}$ for $n_{\rm e} \sim 10^5\,{\rm cm^{-3}}$.  

\section{Accretion}\label{sect:accretion}

Different methods have been described in the literature for estimating the mass accretion
rate ($\dot{M}_{\rm acc}$) from the properties of emission lines or continuum emission.
The various accretion diagnostics probe several physical regions, e.g. 
the excess continuum that veils the 
absorption features of classical T Tauri stars is ascribed to accretion shocks 
\citep{Calvet98.1} and the high-excitation HeI emission to post-shock regions 
\citep{Beristain01.1}. 
Hydrogen lines from the Paschen and Balmer series are produced in accretion flows, with H$\alpha$, 
having the largest optical depth, originating in their outermost parts 
\citep{Hartmann94.1, Muzerolle98.3}.

Here, we use two approaches for measuring the mass accretion rate of \futau:
We make use of empirical relations 
with emission line fluxes and luminosities (Sect.~\ref{subsubsect:accretion_mdot_lines})
and we model the continuum excess (Sect.~\ref{subsubsect:accretion_mdot_cont}).  

\subsection{Emission line analysis}\label{subsect:accretion_lines}

Using IRAF we have measured equivalent widths and line fluxes of the dominant 
emission lines in the spectrum in the same manner as described in 
Sect.~\ref{sect:outflows} for the case of the FELs. 

The by far strongest line in the X-Shooter spectrum of \futau is H$\alpha$.
We measure an equivalent width of $93 \pm 5$\,\AA. Previous H$\alpha$ measurements 
of \futau reported in the literature, all based on lower-resolution spectra, range between 
$93$ and $155$\,\AA~ \citep{Luhman09.2, Stelzer10.0, Scholz12.0}. Considering that
low-resolution observations tend to underestimate line equivalent widths, this points at 
only modest variability. 
The Balmer series can be reliably identified up to $n=16$. No lines of the Paschen or
Brackett series are detected. 

\begin{figure*}
\begin{center}
\parbox{18cm}{
\parbox{6.0cm}{
\includegraphics[width=6.0cm]{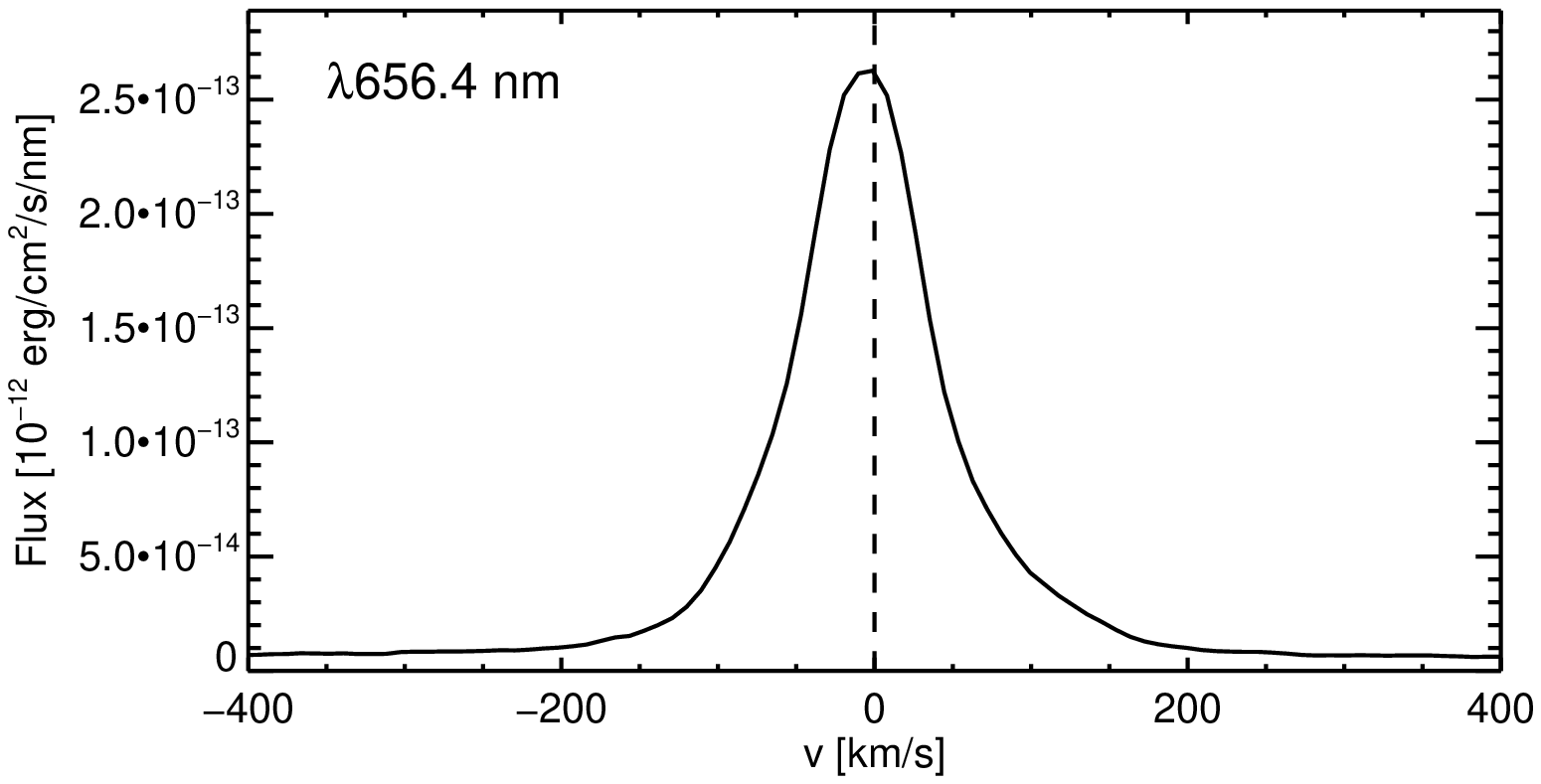}
}
\parbox{6.0cm}{
\includegraphics[width=6.0cm]{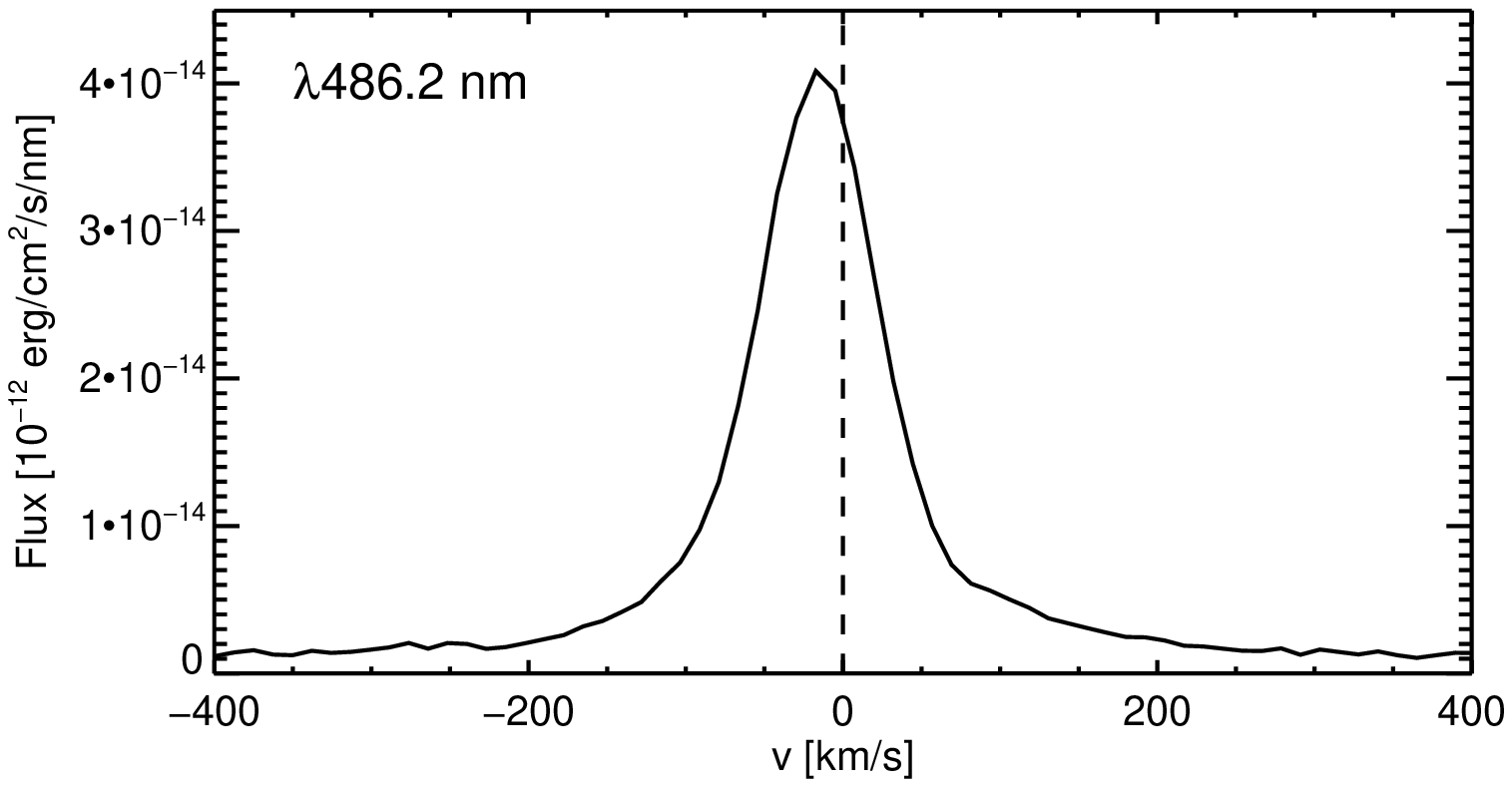}
}
\parbox{6.0cm}{
\includegraphics[width=6.0cm]{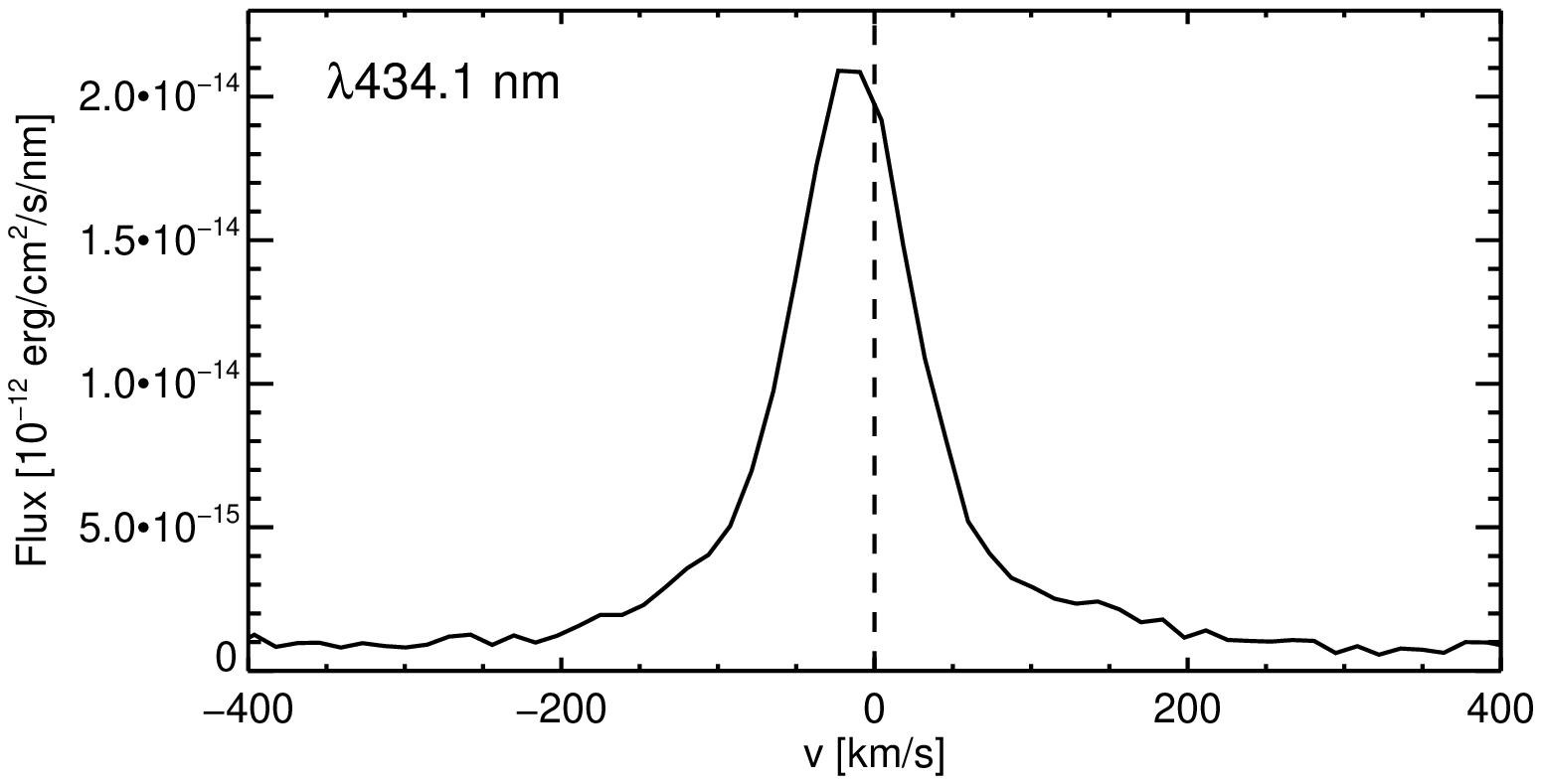}
}
}
\parbox{18cm}{
\parbox{6.0cm}{
\includegraphics[width=6.0cm]{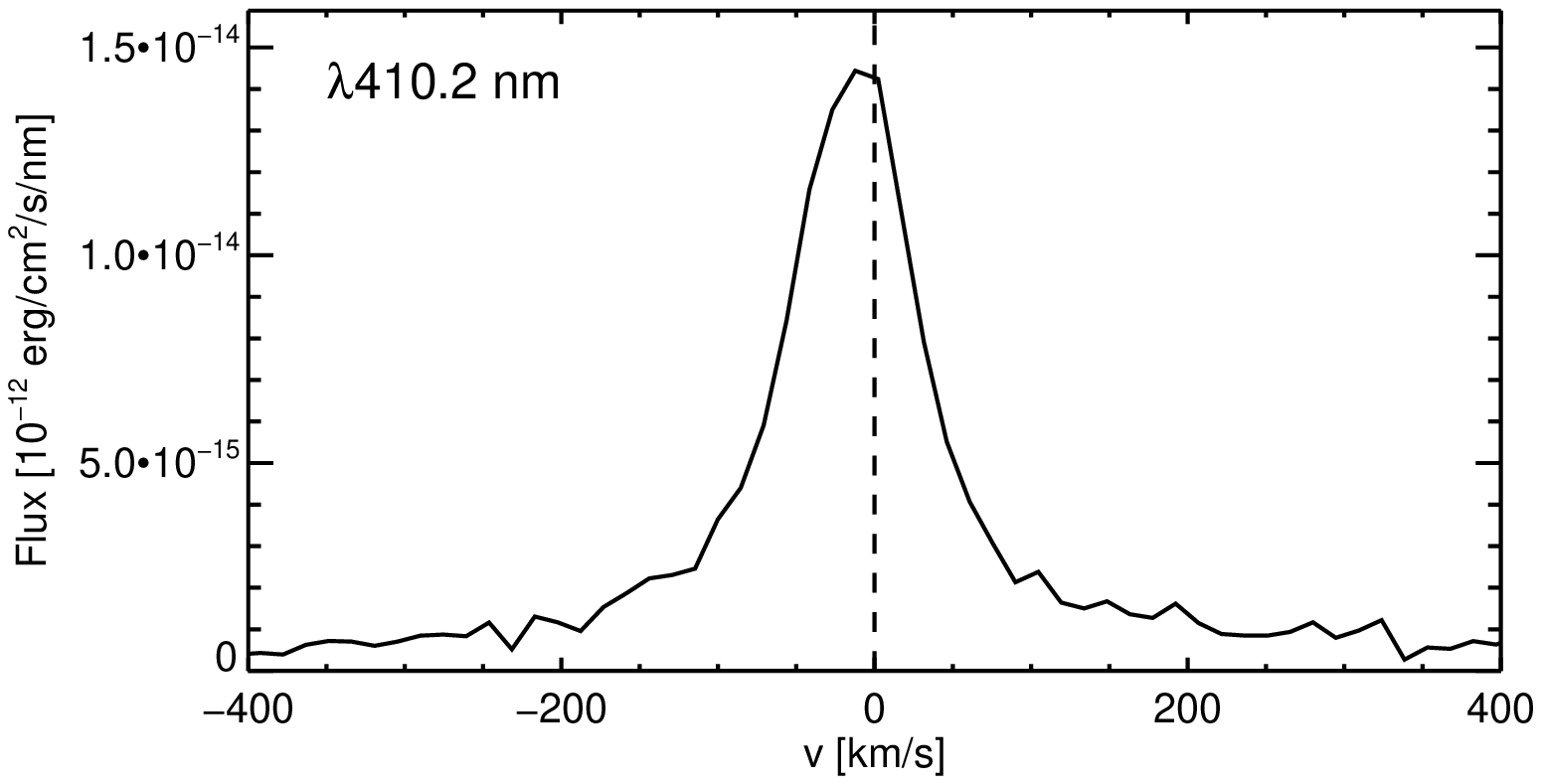}
}
\parbox{6.0cm}{
\includegraphics[width=6.0cm]{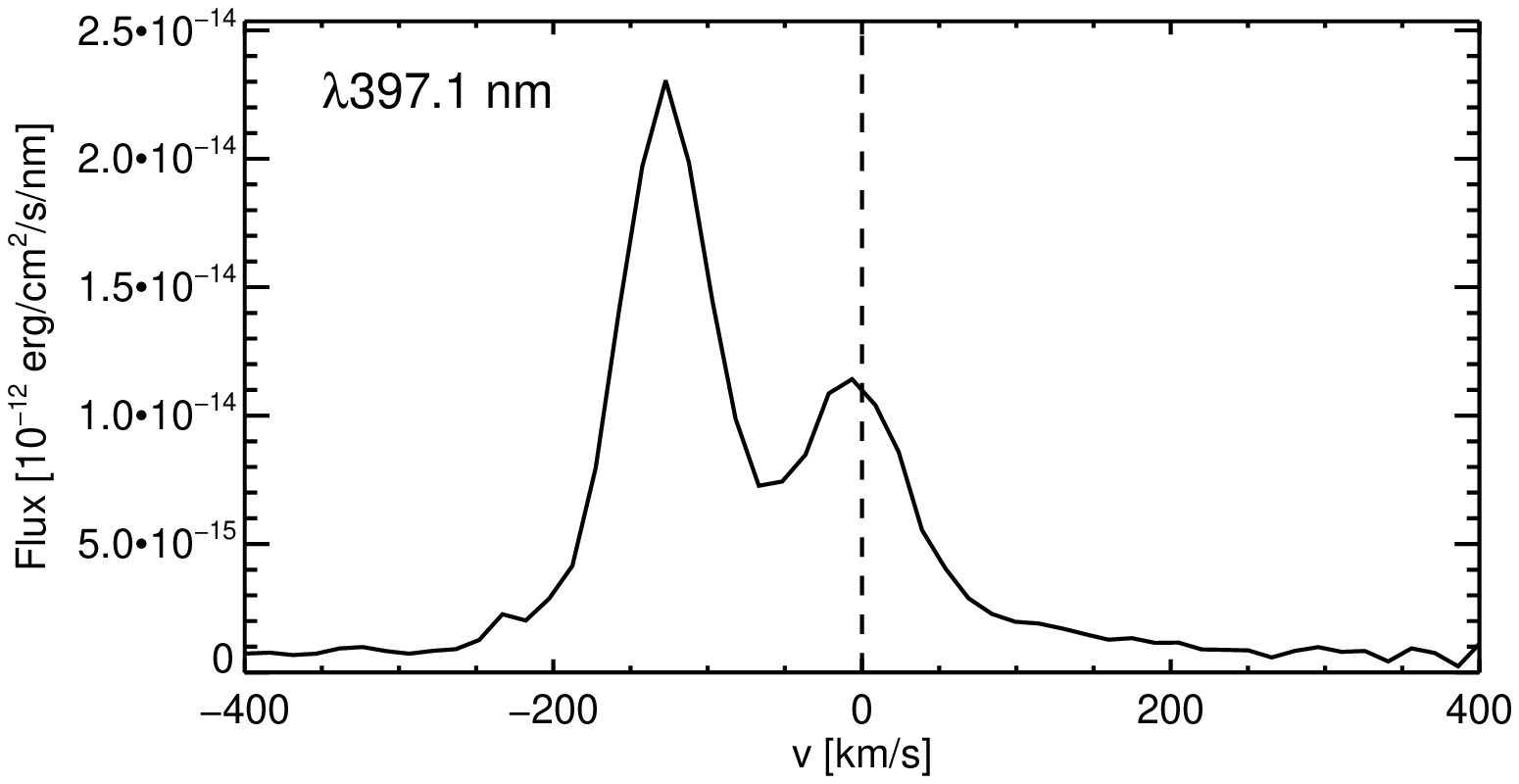}
}
\parbox{6.0cm}{
\includegraphics[width=6.0cm]{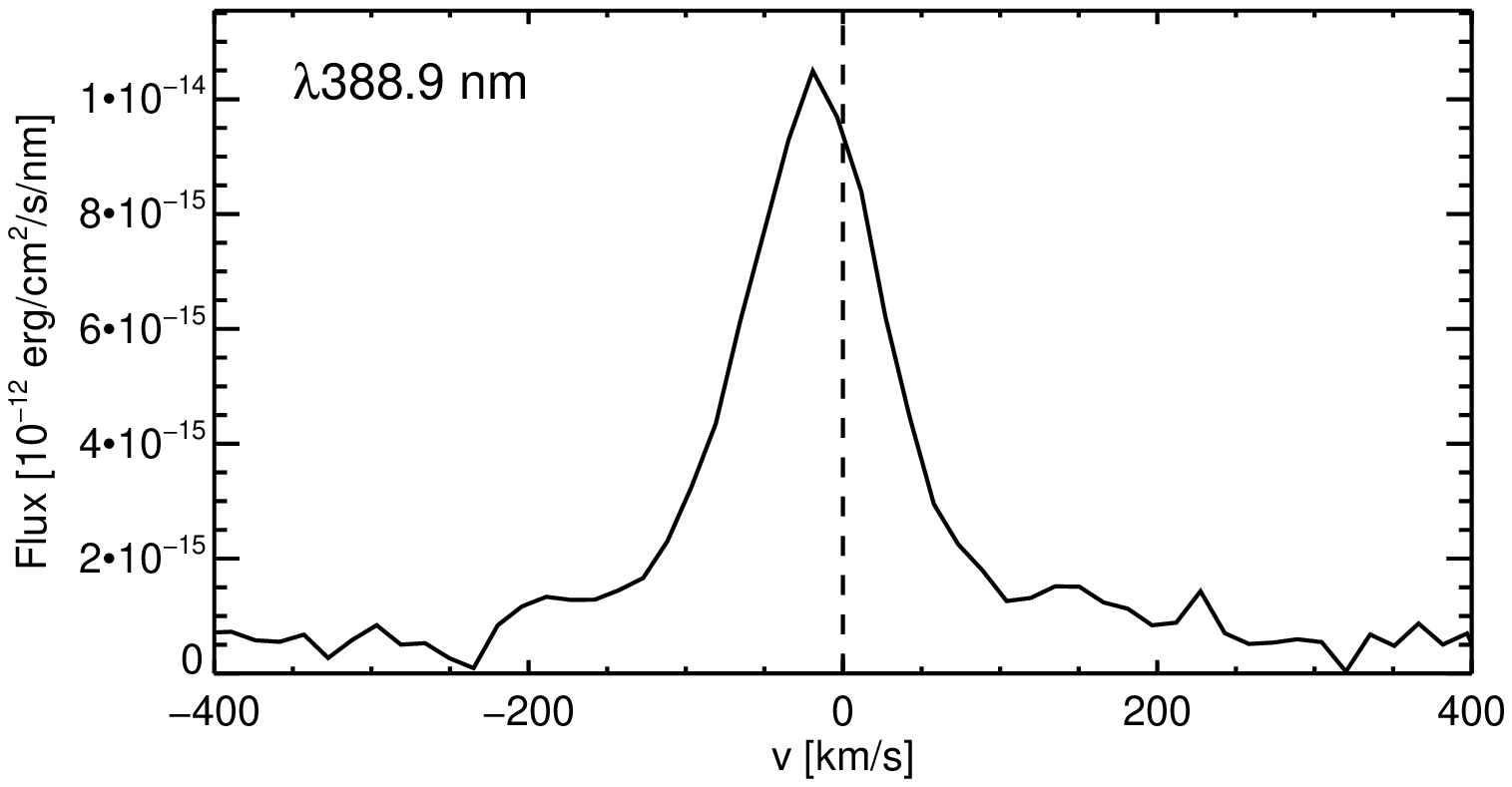}
}
}
\parbox{18cm}{
\parbox{6.0cm}{
\includegraphics[width=6.0cm]{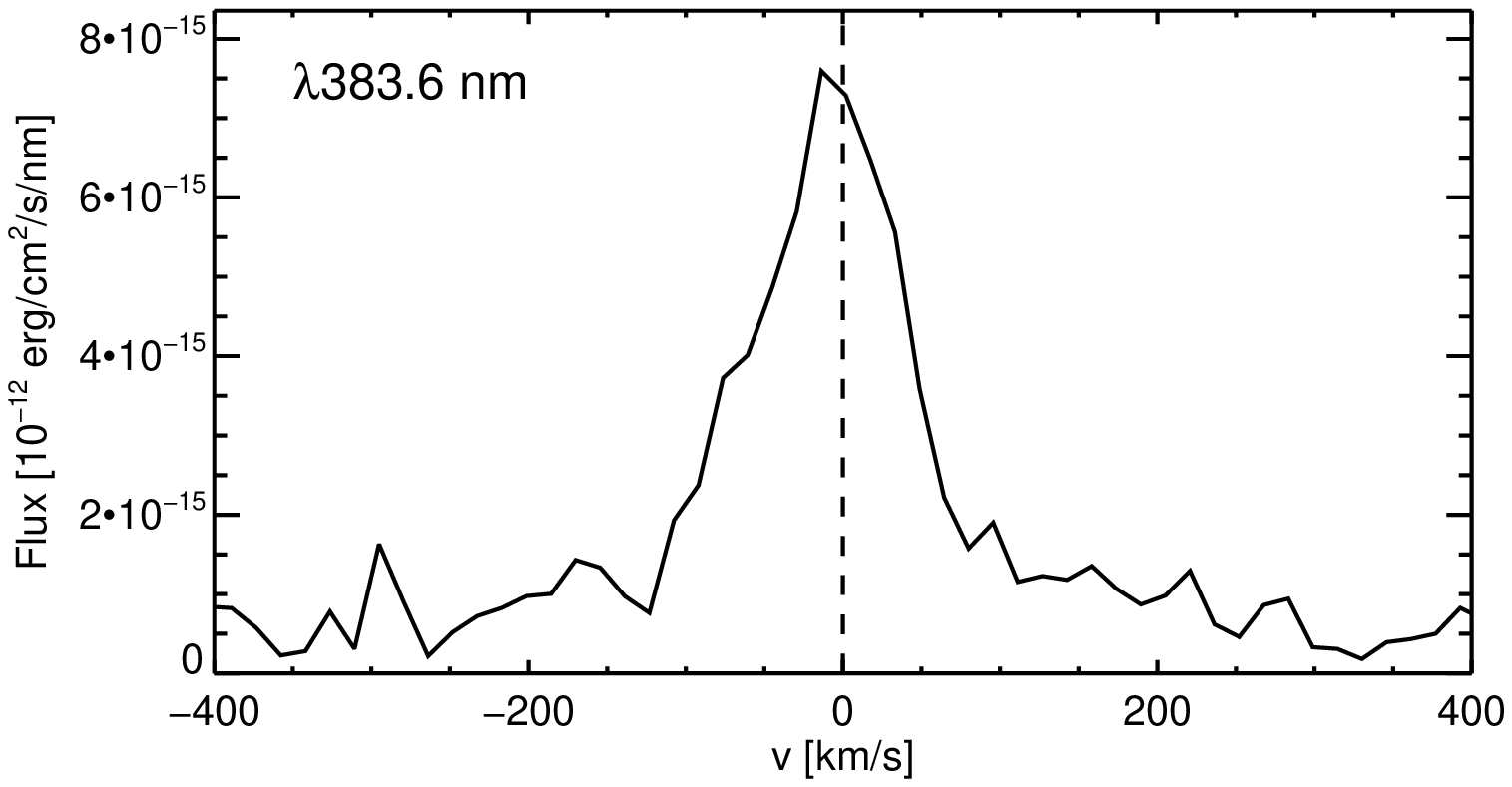}
}
\parbox{6.0cm}{
\includegraphics[width=6.0cm]{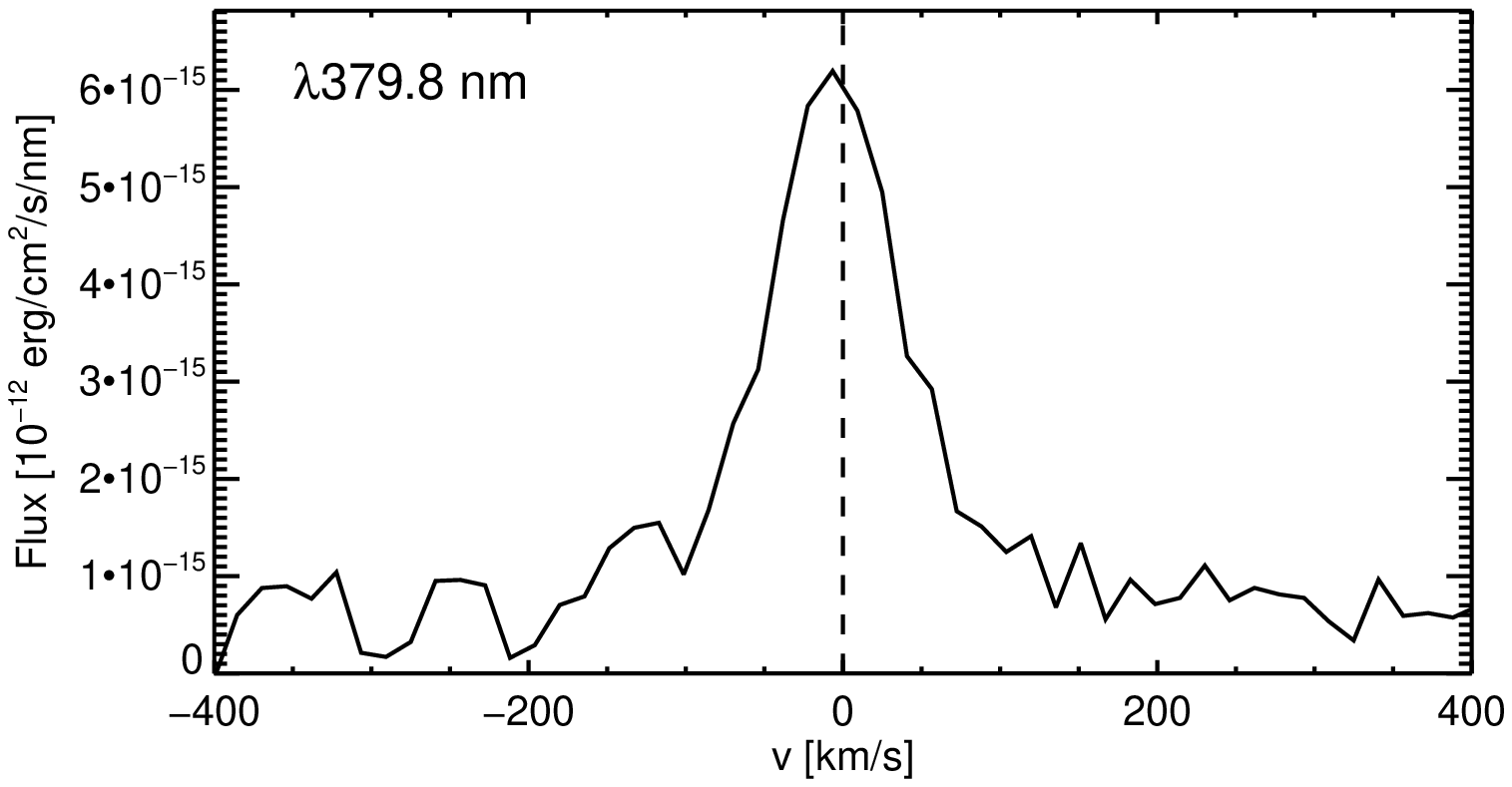}
}
\parbox{6.0cm}{
\includegraphics[width=6.0cm]{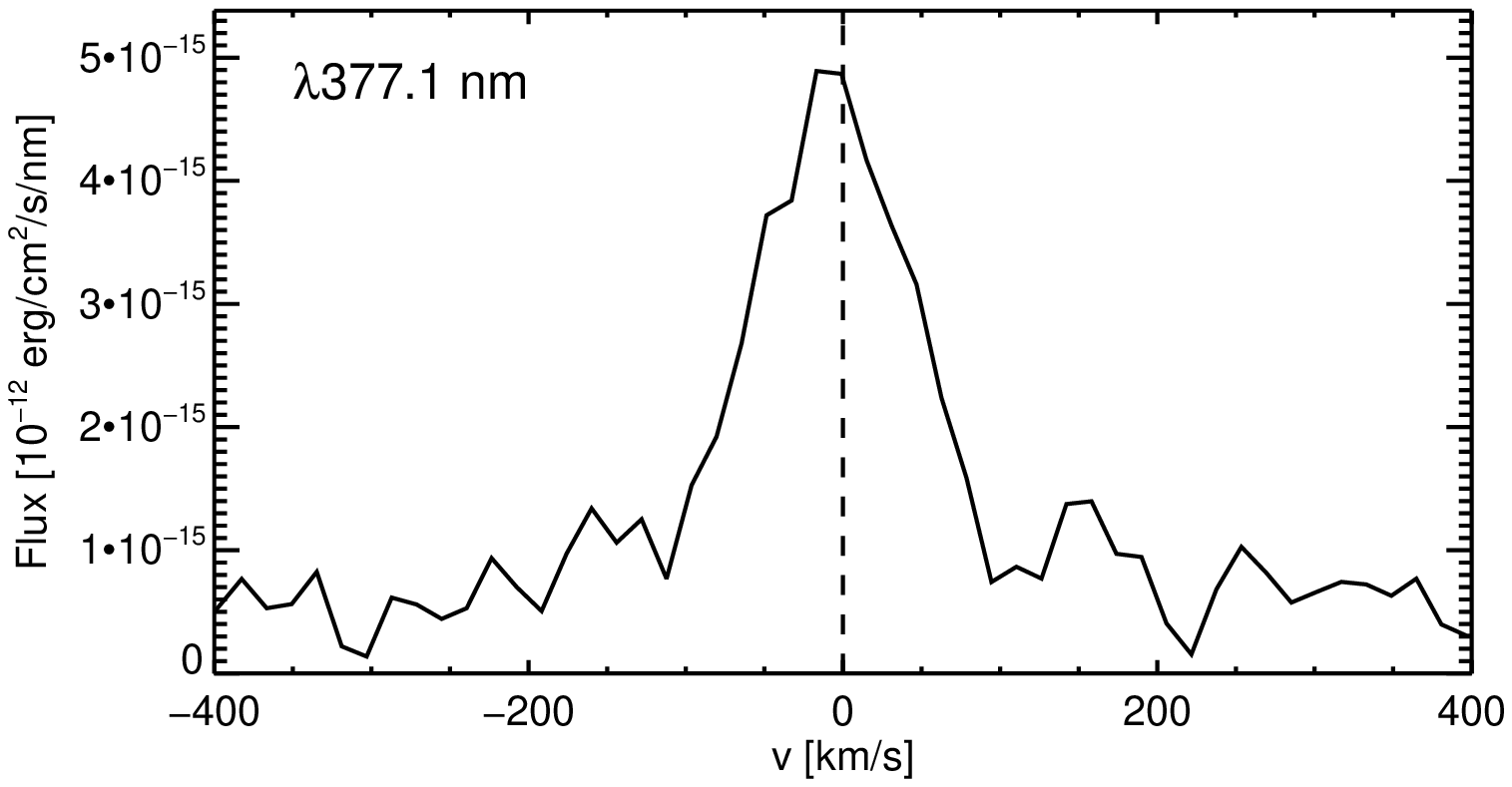}
}
}
\caption{Line profiles for the lower Balmer series of \futau. 
The dashed vertical line represents the
expected line center correcting for the barycenter motion and the stellar $RV$ computed from
absorption lines.}
\label{fig:spec_halpha}
\end{center}
\end{figure*}

Some of the Balmer line profiles are displayed in Fig.~\ref{fig:spec_halpha}. 
All Balmer lines including H$\alpha$ are single-peaked. 
The vertical lines in Fig.~\ref{fig:spec_halpha} represent the expected line center corrected
for barycenter motion and the systemic $RV$ of \futau derived in Sect.~\ref{subsect:results_rv}. 
While the H$\alpha$ profile is nearly symmetric the other low-n Balmer lines show
small asymmetries with a flux deficiency on the red side of the profile. This  
might be interpreted as red absorption, a typical signature of infall \citep{Hartmann94.1}. 

\subsection{Mass accretion rates}\label{subsect:accretion_mdot}

\subsubsection{Emission line diagnostics}\label{subsubsect:accretion_mdot_lines}

The published calibrations between emission line strength and mass accretion rate come in two
flavors. The first consists in relations between line luminosity ($L_{\rm line}$) and 
accretion luminosity ($L_{\rm acc}$) of the type
\begin{equation}
\log{(L_{\rm acc}/L_\odot)} = a \cdot \log{(L_{\rm line}/L_\odot)} + b
\label{eq:lacc_lline}
\end{equation}
This can be converted to accretion rate according to 
\begin{equation}
\dot{M}_{\rm acc} = 1.25 \cdot \frac{L_{\rm acc} \cdot R_*}{G \cdot M_*}
\label{eq:mdot_lacc}
\end{equation}
where the numerical factor derives from the assumption for the inner disk radius,
$R_{\rm in} = 5 R_*$ \citep{Gullbring98.1}.  
The second approach relates the line surface flux ($F_{\rm s,line}$) 
directly to the accretion rate in the form of 
\begin{equation}
\log{\dot{M}_{\rm acc}} = c \cdot \log{F_{\rm s,line}} + d
\label{eq:mdot_fsurf}
\end{equation}
The coefficients $a$, $b$, $c$ and $d$ have been derived for an increasing sample of 
YSOs, however, so far without consensus between the various studies.

\begin{figure*}
\begin{center}
\includegraphics[width=18cm]{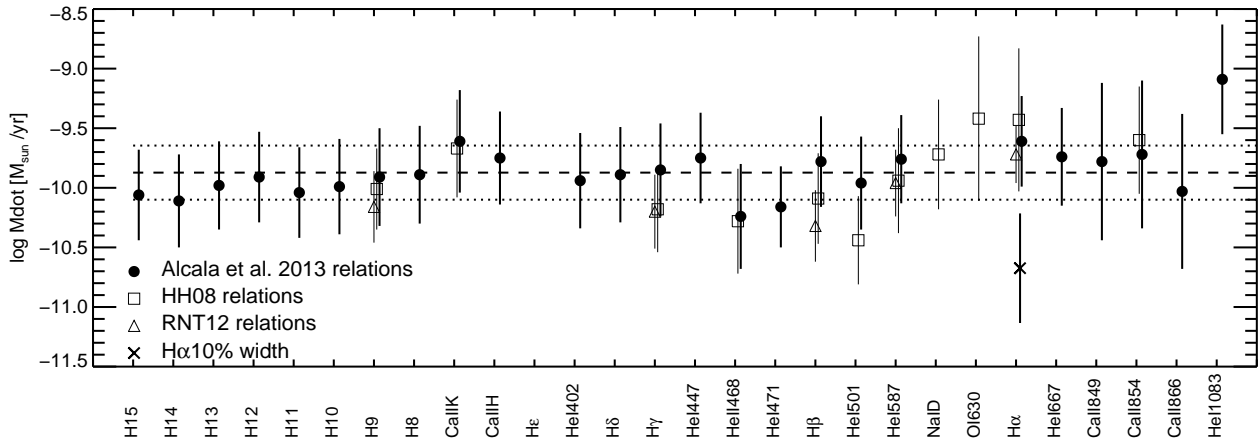}
\caption{Mass accretion rate for \futau derived from various emission line fluxes and
luminosities and from the $10$\,\% width of H$\alpha$ using different calibrations provided
in the literature as labeled in the figure. Dashed and dotted lines represent the mean and
standard deviation for the relations from A13 using $24$ emission lines.}
\label{fig:mdots}
\end{center}
\end{figure*}

To estimate $\dot{M}_{\rm acc}$ for \futau  
we have used the published relations presented by \cite{Herczeg08.1}, \cite{Mohanty05.1}, 
Rigliaco et al. (2012, A\&A in press; hereafter RNT12) 
as well as new calibrations by Alcal\'a et al.,(in prep).
The latter ones are based on an X-Shooter sample of $36$ YSOs from the Lupus star forming region
and represent the largest homogeneous analysis of this kind performed so far.  
The distance used to convert line fluxes to 
luminosities is that of the Taurus star 
forming complex, $140$\,pc. The mass and radius of \futau required for 
evaluating Eqs.~\ref{eq:mdot_lacc} and~\ref{eq:mdot_fsurf} 
are given in Sect.~\ref{subsect:results_params}. 
The derived mass accretion rates for all detected emission lines in common with the 
relations presented in the literature are shown in Fig.~\ref{fig:mdots}. 
Each literature source is marked with a different plotting symbol. 
Some values that deviate strongly from the bulk of the measurements are not shown for
clarity but discussed below. 
The error bars represent again the uncertainty in the reddening with upper and lower
bounds corresponding to $A_{\rm V} = 0$ and $A_{\rm V} = 1$\,mag, respectively.

We have calculated the mass accretion rate for all emission lines detected in the
X-Shooter spectrum of \futau that have been calibrated in the literature. 
However, we have computed the average of $\dot{M}_{\rm acc}$ only from those lines 
that we consider the most reliable accretion diagnostics. 
In particular, we excluded the \ion{Ca}{ii}\,IRT lines and \ion{Na}{i}\,D, for which
strongly discrepant accretion rates are found with different calibrations. Some of the
values for $\dot{M}_{\rm acc}$ that we have derived for these lines are outside the range of
values shown in Fig.~\ref{fig:mdots}. For a detailed
investigation of the reliability of individual emission lines as accretion diagnostic see 
e.g. Alcala et al. (in prep.). We also   
did not take into account [\ion{O}{i}]$\lambda\,630$\,nm and H$\alpha$ that are often 
affected by winds in YSOs (see Sect.~\ref{sect:outflows}). \ion{Ca}{ii}\,K and H$\epsilon$ are also
excluded because these lines are partially blended in our spectrum. 
The average obtained from the remaining lines shown in Fig.~\ref{fig:mdots}
is 
$\langle \log{\dot{M}_{\rm acc}} \rangle {\rm [M_\odot/yr]} = -9.9 \pm 0.2$ for $24$ lines from Alcala et al.,
$\langle \log{\dot{M}_{\rm acc}} \rangle {\rm [M_\odot/yr]} = -9.8 \pm 0.2$ for $6$ lines from RNT12, and 
$\langle \log{\dot{M}_{\rm acc}} \rangle {\rm [M_\odot/yr]} = -10.0 \pm 0.3$ for $11$ lines from HH08. 
Here the uncertainties represent the standard deviation of all measurements. 
The mean and standard deviation of the calibrations by Alcala et al. 
are overplotted in Fig.~\ref{fig:mdots}.
The influence of the uncertainties of mass and distance on this result is examined 
in Sect.~\ref{sect:discussion}. We anticipate here that they make $\log{\dot{M}_{\rm acc}}$ change  
only marginally. Similarly, a smaller value for the inner disk radius, 
e.g. $R_{\rm in} = 2\,R_\odot$, would yield a mass accretion rate that is
marginally compatible with the value given above. 

We have also derived the mass accretion rate from the 
H$\alpha$ $10$\,\% width. The measured full width half maximum at $10$\,\% of the peak height 
is $228$\,km/s. For the calibration provided by \cite{Natta04.2} this corresponds to a
mass accretion rate of $\log{\dot{M}_{\rm acc}}\,{\rm [M_\odot/yr]} = -10.7 \pm 0.5$. 
This value, also shown in Fig.~\ref{fig:mdots}, is almost one order of magnitude
lower than the mean value derived from the relations presented 
by Alcala et al.  

\subsubsection{Continuum excess}\label{subsubsect:accretion_mdot_cont}

\begin{figure}
\begin{center}
\includegraphics[width=9cm]{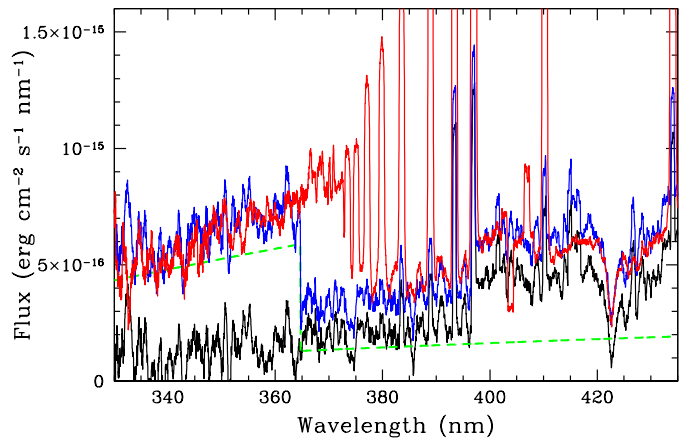}
\caption{Balmer jump region in the spectrum of \futau, 
heavily smoothed to reduce the errors on the continuum (red), 
the Class\,III template SO\,925 (black), the excess emission from the accretion 
slab model (dashed green) and the resulting modelled emission, i.e. template plus excess emission, 
in blue. All spectra are normalized in the region $700-720$\,nm.
Note that the observed emission in the range $346.6$\,nm to $\sim 370$\,nm 
is dominated by unresolved 
Balmer lines, which are not included in the slab model.}
\label{fig:slab}
\end{center}
\end{figure}

The accretion luminosity can be estimated directly by measuring the continuum 
emission in excess of the photospheric one. This is done by comparison to a
non-accreting template star of the same spectral type.  
Unfortunately, the Class III Par-Lup3-1 that we have used as
spectral template in Sect.~\ref{subsect:results_spt} is too noisy to be used for this purpose. 
Fig.~\ref{fig:slab} shows the blue part of the spectrum of \futau compared to that of 
another young Class\,III star of similar (albeit slightly earlier) spectral type,  
SO\,925 (M5.5; see RNT12 and Manara et al. 2013) normalized in the 
region $700-720$\,nm. 
\futau shows a clear excess emission in the Balmer continuum 
and a relatively small Balmer jump, defined as the ratio of the fluxes observed in the 
continuum at $360$\,nm and at $420$\,nm, of $\sim 1.2$. 
The Balmer limit is at $346.6$\,nm but due to the blending of the higher Balmer lines
it appears shifted to $370$\,nm. 

We model the excess emission as due to a slab of hydrogen of given electron density, 
temperature and length, \cite[see e.g.,][]{Valenti93.0}; HH08; RNT12. 
 and obtain an estimate of the excess continuum luminosity of 
 $L_{\rm acc} \sim 1.3 \times 10^{-4}\,{\rm L_\odot}$. 
 Fig.~\ref{fig:slab} shows the best-fitting model in blue. It reproduces quite well the 
 observed Balmer continumm, the Balmer jump and the profile of the \ion{Ca}{ii} line at 
 $420$\,nm. Veiling of 
 photospheric lines at longer wavelengths is very weak as shown in 
Sect.~\ref{subsect:results_lithium}.  
The uncertainties of $L_{\rm acc}$ 
are quite large ($50$\,\% at least),  due to the noise of the spectra 
of both \futau and the Class\,III template and due to the mismatch between the spectral 
type of \futau and that of the template.
For $M = 0.08\,{\rm M_\odot}$, $R_* = 1.38\,{\rm R_\odot}$ 
as determined in Sect.~\ref{subsect:results_params}
the corresponding mass accretion rate from Eq.~\ref{eq:mdot_lacc} is  
$\log{\dot{M}_{\rm acc}}\,{\rm [M_\odot/yr]} \sim -10.1$, 
in reasonably good agreement with the estimates from the emission lines presented in 
Sect.~\ref{subsubsect:accretion_mdot_lines}. 

\begin{table}\begin{center}
\caption{Equivalent widths and fluxes for emission lines.}
\label{tab:lines}
\begin{tabular}{lrcc} \hline
Line  & $\lambda_0$ & $EW$ & $\log{f_{\rm line}}$ \\
      & [nm]      & [nm]      & [${\rm erg/cm^2/s/nm}$] \\ \hline
    HeI 1083 & $  1083.038$ & $ -0.150 \pm   0.026$ & $  -13.73 ^{+ 0.06}_{- 0.07 }$ \\
    CaII 866 & $   866.381$ & $ -0.017 \pm   0.003$ & $  -15.11 ^{+ 0.10}_{- 0.11 }$ \\
    CaII 854 & $   854.347$ & $ -0.076 \pm   0.015$ & $  -14.61 ^{+ 0.10}_{- 0.12 }$ \\
    CaII 849 & $   849.936$ & $ -0.042 \pm   0.010$ & $  -14.86 ^{+ 0.10}_{- 0.12 }$ \\
     HeI 667 & $   667.928$ & $ -0.089 \pm   0.014$ & $  -15.40 ^{+ 0.18}_{- 0.14 }$ \\
   H$\alpha$ & $   656.374$ & $ -9.766 \pm   0.709$ & $  -13.18 ^{+ 0.17}_{- 0.16 }$ \\
     ${\rm [OI]}$630 & $   630.093$ & $ -0.314 \pm   0.039$ & $  -15.09 ^{+ 0.15}_{- 0.19 }$ \\
      NaI D1 & $   589.697$ & $ -0.221 \pm   0.044$ & $  -15.51 ^{+ 0.18}_{- 0.21 }$ \\
      NaI D2 & $   589.096$ & $ -0.372 \pm   0.067$ & $  -15.27 ^{+ 0.18}_{- 0.20 }$ \\
     HeI 587 & $   587.670$ & $ -0.581 \pm   0.065$ & $  -15.01 ^{+ 0.20}_{- 0.18 }$ \\
     HeI 501 & $   501.625$ & $ -0.071 \pm   0.016$ & $  -16.07 ^{+ 0.18}_{- 0.24 }$ \\
    H$\beta$ & $   486.184$ & $ -6.360 \pm   1.765$ & $  -14.14 ^{+ 0.22}_{- 0.23 }$ \\
     HeI 471 & $   471.397$ & $ -0.031 \pm   0.009$ & $  -16.50 ^{+ 0.15}_{- 0.24 }$ \\
    HeII 468 & $   468.633$ & $ -0.048 \pm   0.016$ & $  -16.30 ^{+ 0.20}_{- 0.23 }$ \\
     HeI 447 & $   447.216$ & $ -0.470 \pm   0.164$ & $  -15.47 ^{+ 0.24}_{- 0.23 }$ \\
   H$\gamma$ & $   434.096$ & $ -6.334 \pm   2.929$ & $  -14.44 ^{+ 0.24}_{- 0.23 }$ \\
   H$\delta$ & $   410.221$ & $ -4.151 \pm   2.327$ & $  -14.64 ^{+ 0.22}_{- 0.26 }$ \\
     HeI 402 & $   402.687$ & $ -0.238 \pm   0.120$ & $  -15.87 ^{+ 0.25}_{- 0.26 }$ \\
 H$\epsilon$ & $   397.068$ & $ -1.767 \pm   1.078$ & $  -14.89 ^{+ 0.25}_{- 0.26 }$ \\
      CaII H & $   396.898$ & $ -5.297 \pm   3.764$ & $  -14.63 ^{+ 0.28}_{- 0.20 }$ \\
      CaII K & $   393.419$ & $ -5.981 \pm   2.717$ & $  -14.52 ^{+ 0.25}_{- 0.25 }$ \\
          H8 & $   388.944$ & $ -4.875 \pm   3.823$ & $  -14.78 ^{+ 0.25}_{- 0.25 }$ \\
          H9 & $   383.588$ & $ -2.392 \pm   1.852$ & $  -14.98 ^{+ 0.26}_{- 0.24 }$ \\
         H10 & $   379.845$ & $ -0.975 \pm   0.447$ & $  -15.15 ^{+ 0.24}_{- 0.26 }$ \\
         H11 & $   377.113$ & $ -0.855 \pm   0.512$ & $  -15.25 ^{+ 0.22}_{- 0.27 }$ \\
         H12 & $   375.071$ & $ -1.230 \pm   0.684$ & $  -15.28 ^{+ 0.24}_{- 0.26 }$ \\
         H13 & $   373.484$ & $ -0.658 \pm   0.422$ & $  -15.44 ^{+ 0.18}_{- 0.25 }$ \\
         H14 & $   372.242$ & $ -0.384 \pm   0.250$ & $  -15.66 ^{+ 0.19}_{- 0.27 }$ \\
         H15 & $   371.232$ & $ -0.377 \pm   0.276$ & $  -15.64 ^{+ 0.25}_{- 0.28 }$ \\
         H16 & $   370.432$ & $ -0.105 \pm   0.057$ & $  -16.04 ^{+ 0.22}_{- 0.25 }$ \\
\hline
\end{tabular}
\end{center}\end{table}

\section{Discussion}\label{sect:discussion}

We have analysed a broad-band ($300-2480$\,nm) medium-resolution ($R \sim 5000 - 9000$) spectrum
of \futau obtained with X-Shooter. On the basis of these data we have presented 
the first measurements
of gravity, radial velocity, rotational velocity and lithium content as well as the first
detection of outflow activity in \futau. Moreover, this spectrum provides a large number 
of accretion diagnostics which we have employed 
for a detailed evaluation of the mass accretion rate.

The motivation for our spectroscopic study was the search for an explanation of the
(apparent) overluminosity of \futau with respect to the predictions of evolutionary models
for its age and mass. This requires precise measurements of the fundamental stellar parameters
($T_{\rm eff}$, $L_{\rm bol}$, $R_*$, $\log{g}$) as well as a study of other phenomena such as 
accretion and magnetic activity that may lead to a wrong estimate of those parameters. 
We have discussed several possible factors that may influence the position of \futau in 
the HR diagram in our previous papers \citep{Stelzer10.0, Scholz12.0} and we resume this
discussion here, adding the wealth of information provided by the X-Shooter spectrum.

\subsection{Accretion and outflow in \futau}\label{sect:disc_inout}

In our detailed investigation of mass accretion in \futau  
we made use of an unprecedented large number
of empirical relations between the luminosity of individual emission lines and the accretion
luminosity derived from a comprehensive X-Shooter study of accretors in the Lupus clouds
(Alcala et al., in prep.). 
These relations comprise emission features collected with all three arms of the 
X-Shooter spectrograph. As compared to similar relations presented in the literature, we consider 
them the most reliable calibrations for our observation
of \futau given that they were derived with the same instrument. 
A mean value of $\langle \log{\dot{M}_{\rm acc}} \rangle {\rm [M_\odot/yr]} = -9.9 \pm 0.2$ 
is found from a total of $24$ emission lines. 
The accretion rate obtained from the H$\alpha 10$\,\% width 
($\log{\dot{M}_{\rm acc,10\%}}\,{\rm [M_\odot/yr]} = -10.7 \pm 0.5$) 
is lower by almost one order of magnitude. The discrepancy of 
$\log{\dot{M}_{\rm acc,10\%}}$ with respect to other accretion diagnostics supports previous
notions of the poor reliability of this tracer (e.g., Costigan et al. 2012, MNRAS in press). 
On the other hand,  
the accretion rate from the H$\alpha 10$\,\% width as measured in the X-Shooter spectrum
is also much lower than the value measured with the same diagnostics from a previous Gemini 
spectrum \citep[$\log{\dot{M}_{\rm acc,10\%}}\,{\rm [M_\odot/yr]} = -9.5$;][]{Stelzer10.0}. 
The low velocities in the new H$\alpha$ profile are not sufficient to explain the temperature
of the X-ray emitting plasma observed by \cite{Stelzer10.0} as originating in an accretion shock. 

We have presented the first evidence for outflow activity in \futau in the form of 
several forbidden emission lines. The outflow of \futau has recently also been
detected at millimeter wavelengths (Monin et al., A\&A subm.).  
In this paper, we have used 
the [OI] $\lambda 630.0$\,nm  and [SII] $\lambda 673.1$\,nm lines detected in
the X-Shooter spectrum to derive the 
mass loss rate of the outflow using the approach presented by \cite{Hartigan95.1}. 
We trust the result from [SII] $\lambda 673.1$\,nm more than that for 
[OI] $\lambda 630.0$\,nm because the former has a weaker dependence on the unknown electron density. 
An additional uncertainty is introduced by the fact that we have only a lower limit
on the disk inclination angle that determines the transverse outflow velocity. Assuming 
$i = 53^\circ$ and $n_{\rm e} \sim 10^{-3}\,{\rm cm^{-2}}$ we derive from 
the [SII] $\lambda 673.1$\,nm flux  
a mass outflow rate of $\log{\dot{M}}_{\rm out}\,{\rm [M_\odot/yr]} \sim -10.4$.  
Combining this with the contemporaneous measurement of the mass accretion rate 
we find an outflow-to-inflow rate $\dot{M}_{\rm out}/\dot{M}_{\rm acc} \sim 0.3$. 
This value for the mass outflow rate of \futau could be both a lower limit 
(given that $n_{\rm e}$ is an upper limit) or an 
upper limit (given that $i$ is a lower limit); 
see Sect.~\ref{sect:outflows} for details. 
Very few measurements of $\dot{M}_{\rm out}/\dot{M}_{\rm acc}$ have been presented in the
literature for VLM objects. While some of these observations suggest that mass inflow and
outflow rate are similar in the VLM regime \citep{Bacciotti11.0, Whelan09.0}, others 
point at results close to the `canonical' value for higher-mass cTTS of 
$\dot{M}_{\rm out}/\dot{M}_{\rm acc} \sim 0.1...0.01$ \citep{Joergens12.0}.
A final conclusion on the comparison of the mass outflow to inflow ratio between 
BDs and cTTSs and the comparison of observations to the predictions of jet-launching
models \citep{Cabrit09.0} is premature considering the large uncertainties associated 
with the observations 
of $\dot{M}_{\rm out}$ and $\dot{M}_{\rm acc}$ in VLM objects and the 
inhomogeneous approach by different authors for calculating these numbers. 

Signs for in- and outflows of \futau are also evident from the 
line profiles and line shifts observed in the X-Shooter spectrum. 
The modest deficiency of flux on the red side of the low-n Balmer lines is in qualitative agreement
with the outcome of magnetospheric accretion models for high inclination. 
The calculations by \cite{Hartmann94.1} predict 
a lack of flux in the red wing of optically thin emission lines that increases with decreasing 
inclination angle due to occultation of the receeding part of the accretion flow by the disk. 
The strongest forbidden lines show a similar blueshift of $\approx -10$\,km/s with respect 
to the stellar motion. This
emission likely represents the approaching lobe of the jet, while the 
receeding (redshifted) part of the outflow may be occulted by the accretion disk.
These are plausible interpretations which are consistent with our range for the 
disk inclination angle (derived from $v \sin{i}$, rotation period and stellar radius) 
that excludes a pole-on view.

\subsection{\futau in the HR diagram}\label{subsect:disc_hrd}

\subsubsection{Extinction}\label{subsubsect:disc_av}

A wrong extinction estimate could, in principle, misplace the object in the HR diagram
in the vertical direction. 
We have determined the spectral type of \futau to M6.5 $\pm$ 0.4 and the optical extinction
to $A_{\rm V} = 0.5 \pm 0.5$\,mag. This is slightly hotter and less absorbed than found by
\cite{Luhman09.1}, and results in a higher effective temperature ($T_{\rm eff} = 2940$\,K) 
and smaller bolometric luminosity ($L_{\rm bol} = 0.13\,{\rm L_\odot}$)
and radius ($R_* = 1.38\,{\rm R_\odot}$).
The smaller extinction 
moves the object vertically down in the HR diagram by less than a factor two.
This is by far not sufficient for resolving the luminosity problem of \futau.

\subsubsection{Accretion}\label{subsubsect:disc_accretion}

A possible explanation for the overluminosity of \futau with respect to the youngest
isochrones of evolutionary models could be an overestimate of the bolometric luminosity
due to a substantial contribution of accretion luminosity. 
In \cite{Scholz12.0} we have argued that this scenario is unlikely because of the 
relatively low temperature derived for the accretion hot spots. 
It was also pointed out that a significantly lower value for $L_{\rm bol}$,
when combined with the X-ray data presented by \cite{Stelzer10.0}, would yield
an exceptionally high X-ray to bolometric flux ratio. 
The X-Shooter spectrum has now allowed us to directly compute the accretion luminosity
by modeling of the Balmer continuum. We found that accretion makes up for only 
$0.1$\,\% of the bolometric luminosity of \futau. Therefore, accretion luminosity can not
explain the overluminosity of \futau. This is also corroborated by the absence of
strong veiling in the X-Shooter spectrum. 

Accretion possibly plays a role in another puzzling property of \futau, its weak lithium
absorption (see discussion in Sect.~\ref{subsect:disc_lithium}).

\subsubsection{Magnetic activity}\label{subsubsect:disc_activity}

An alternative possibility to bring \futau in better agreement 
with evolutionary models would be to 
shift the object horizontally towards higher effective temperature and mass.
It was shown by \cite{Chabrier07.1} and \cite{MacDonald09.0} that the influence
of a strong magnetic field onto convection may provide such an effect. 
In this scenario \futau appears cooler and with larger
radius than expected for its mass. As $T_{\rm eff}$ is obtained from observations without
resorting to evolutionary models while mass is model-dependent, 
this would imply a higher mass for
\futau than the $0.08\,M_\odot$ obtained when extrapolating the object vertically down
to the $1$\,Myr isochrone of \cite{Baraffe98.1}. 

In \cite{Stelzer10.0} we had derived a mass of $\sim 0.2\,{\rm M_\odot}$
for \futau assuming this scenario based on the stellar parameters given by \cite{Luhman09.1}
and we had estimated the rotation period and the magnetic field strength using standard
magnetospheric accretion models. 
These estimates can now be updated making use of the
new values for the stellar parameters derived from the X-Shooter spectrum.

First, the new value for the bolometric luminosity 
corresponds to a mass of $0.15\,{\rm M_\odot}$ and a temperature of $3090$\,K on the 
$1$\,Myr isochrone of \cite{Baraffe98.1}. 
Our new value for the mass accretion rate (see Sect.~\ref{subsubsect:accretion_mdot_lines}) 
was based on a mass of $M = 0.08\,{\rm M_\odot}$.
This would be corrected downward to 
$\log{\langle \dot{M}_{\rm acc} \rangle}\,{\rm [M_\odot/yr]} = -10.1 \pm 0.2$. 
Using these new parameters and 
an inner disk truncation radius of $R_{\rm in} \sim 2...5\,{\rm R_*}$ 
the magnetic field strength predicted by \cite{Koenigl91.1} 
is $B \approx 40...180$\,G. No magnetic field measurement has been performed yet for \futau
that would allow to test these numbers. In any case, this estimate shows that 
the combination of stellar parameters and mass accretion rate observed for \futau 
support the presence of a strong surface field. 
The X-wind model \citep[e.g.,][]{Mohanty08.0}, where the Keplerian angular
velocity is equal to the stellar rotational velocity at the inner disk truncation radius,  
 predicts a stellar rotation period of $P_{\rm rot} \approx 1.4...5.5$\,d 
 for the range of $R_{\rm in}$ given above.  The value observed by 
\cite{Scholz12.0} for the rotation period ($P_{\rm rot} = 4.0 \pm 0.2$\,d) 
is consistent with this prediction. As pointed out
in our previous work this period is rather large for substellar mass according to the
empirical trend between period and mass \citep[e.g., ][]{Scholz05.1}. 

The influence of magnetic activity on the stellar parameters is also expected to have an effect
on the lithium depletion of late-type stars but this is unlikely to be relevant for
\futau as we describe in Sect.~\ref{subsect:disc_lithium}.

\subsubsection{Age}\label{subsubsect:disc_age}

Both \futau and its brown dwarf companion FU\,Tau\,B have the highest absolute
$J$ band magnitude of all Taurus members with which they share the effective temperature,
$T_{\rm eff} \sim 2800$\,K and $T_{\rm eff} \sim 2400$\,K respectively \citep{Scholz12.0}.
This suggests that the pair is coeval but younger than the other VLM objects in Taurus.  
The low gravity of $\log{g} = 3.5 \pm 0.5$ measured from the X-Shooter spectrum 
is a confirmation of the pre-main sequence nature of \futau.  
However, in the standard picture extreme youth is not compatible 
with the weak lithium signature.
Furthermore, there are no signs for an envelope which would be expected if the
object were still in a protostellar phase.

\subsubsection{Binarity}\label{subsubsect:disc_binarity}

The picture could be further complicated if \futau was a close binary. 
The maximum decrease in $L_{\rm bol}$ obtained in this scenario is a factor two
which would make \futau similar to some other VLM objects in Taurus that are
overluminous with respect to the \cite{Baraffe98.1} $1$\,Myr isochrone. 
An equal mass binary with $140$\,AU separation ($\approx 1^{\prime\prime}$ at the distance
of Taurus) would have an $RV$ amplitude of less than $\sim 1$\,km/s but a sub-arcsecond
binary would produce a much larger $RV$ signal. 
Binarity could be responsible for some of the difference between the $RV$ we
measured for \futau and most Taurus members. 
One of the most obvious effects of binarity, besides its influence on the $RV$, 
would be that the rotational velocity has an unknown
contribution from the orbital motion and the measured $v \sin{i}$ is an upper limit. 
The $R \sin{i}$ estimate derived from the rotational velocity and the rotation period
is, in fact, slightly larger than the stellar radius obtained from Stefan-Boltzmann's law
($R \sin{i} = 1.6 \pm 0.5\,{\rm R_\odot}$, $R_* = 1.4\,{\rm R_\odot}$) but 
compatible with each other within the errors. 
The lithium equivalent width may also change in the case of binarity 
but in a way which is hard to predict
without knowledge of the relative brightness of both binary components.

\subsubsection{Distance}\label{subsubsect:disc_distance}

Finally, an obvious way to reduce the bolometric luminosity of \futau would be a smaller
distance. 
This would also lead to a better agreement of the X-ray luminosity of \futau with
other objects of similar mass in Taurus \citep[see][]{Stelzer10.0}. For a smaller distance,
the difference between the mean mass accretion rate obtained from the $24$ emission lines
and the mass accretion rate obtained from the H$\alpha$\,10\,\% width of \futau could 
be reconciled, e.g. a distance of $75$\,pc yields 
$\langle \log{\dot{M}_{\rm acc}} \rangle\,{\rm [M_\odot/yr]} = -10.4 \pm 0.2$ for the mean of 
the emission lines with the stellar parameters from Sect.~\ref{subsect:results_params}. 
The magnetic field would be larger ($B = 90...430$\,G) and the
rotation faster ($P_{\rm rot} = 0.8...2.9$\,d) than predicted for the magnetic activity 
scenario of Sect.~\ref{subsubsect:disc_activity}. 
However, the theoretical predictions for these parameters should be taken as face values and 
not as firm prove for one or the other hypothesis. 

The closer distance poses the problem of 
either 
the existence of a dark cloud apparently associated with the Taurus filaments but 
in reality located in the foreground
or the unlikely projection of an isolated young binary onto a background dark cloud. 
An argument in favor of the latter case comes from the difference of the LSR 
velocity of \futau ($12.5 \pm 2.9$\,km/s) and the value derived for Barnard\,215 
from CO maps \citep[$v_{\rm LSR} \sim 7$\,km/s;][]{Narayanan08.0}. On the other hand, 
this scenario where \futau is in the foreground of Barnard\,215, i.e. far from any signatures 
of star formation, 
puts a question mark on its origin. Ejection from a star forming site seems not a viable
option as for typical ejection speeds of $\sim 1$\,km/s \citep[e.g.][]{Bate03.1} 
\futau cannot have travelled
beyond the (projected) diameter of Barnard\,215 within its lifetime. 
\cite{Luhman09.1} mentioned that the SDSS image shows nebulosity
centered and surrounding FU\,Tau which, when combined with the signatures of youth,  
suggests that the object is associated with the dark cloud. 

The $RV$ of \futau ($22.5 \pm 2.9$\,km/s) 
is marginally consistent with the $RV$ distribution of Taurus and so is its $UVW$ space velocity
(($U,V,W$) = ($-22.09$,$-9.08$,$-10.36$)\,km/s). 
For the proper motion from \cite{Luhman09.1} a closer distance results in a 
space motion with a stronger deviation from the mean Taurus $UVW$ velocities 
(e.g., ($U,V,W$) = ($-21.80$,$-3.98$,$-8.64$\,km/s) for $80$\,pc). However, 
considering the uncertainty of the proper motion
inferred from the different literature sources, the change in space motion associated with
a closer distance seems not significant enough to allow us to draw conclusions. 

\subsection{The lithium problem}\label{subsect:disc_lithium}

The lithium equivalent width measured in the X-Shooter spectrum of \futau is 
$\sim 430$\,m\AA.  
This corresponds to a lithium abundance below A(Li) $\sim 1.9$ according to the curves of growth 
calculated by \cite{ZapateroOsorio02.2}. 
At present, \futau defines the lower envelope in the 
lithium equivalent widths of spectral types M6...M7. 
We have shown that this does not change even if weak veiling, 
consistent with the depth of the TiO absorption bands, is assumed. 
The second lowest value is represented by 
KPNO-Tau\,5, another young BD in Taurus \citep{Barrado04.3}. This object has no
disk \citep{Monin10.1} and is below the canonical threshold for accretors in terms of
its H$\alpha$ equivalent width \citep{Barrado04.3} such that veiling seems unlikely
to be present. 
Lithium
measurements are available for a limited sample of such late-type young objects. Therefore, 
no conclusions can be drawn yet on the typical value at those spectral types. However, 
the large dispersion of lithium abundances is seen also at early-M types and has been
noted in the literature as an unsolved problem. 

Quite a number of parameters not taken into account in standard PMS evolutionary models
may affect the lithium depletion timescale. 
\cite{Eggenberger12.0} described the influence of rotation and disk lifetime on the lithium
depletion of solar-mass stars. 
Interestingly, rotation shifts the evolutionary tracks to slightly lower $T_{\rm eff}$
in the HR diagram.  
Similarly, 
if the disks are longer lived lithium gets more and more rapidly depleted.
The impact of rotation and disk lifetime on lithium depletion is at odds with the
fact that, for a given star, ages derived 
from its lithium abundance tend to be older than those obtained from its position in the 
HR diagram. 
\cite{Yee10.0} explain how the HR diagram ages
and the lithium ages can be brought into qualitative agreement by the inflation of the
radii due to magnetic activity described in Sect.~\ref{subsubsect:disc_activity}. The
cooler effective temperatures associated with the larger radius would imply for a given
star an older age from the HR diagram and a younger lithium depletion age. 
However, all these studies regard solar-type stars and their validity in the VLM regime
can not be taken for granted. Moreover, these arguments should break down for objects as 
young as \futau where no lithium depletion at all is expected from pre-main sequence models. 

For \futau another scenario is more plausible. 
\cite{Baraffe10.0} argue that episodic accretion bursts during the pre-main sequence 
phase strongly decrease the timescale for lithium depletion in low-mass stars.
In particular, they show that an accreting $0.1\,M_\odot$ star may completely deplete its 
initial lithium content within $10$\,Myrs compared to $> 50$\,Myrs required in the non-accreting
case. For suitable values of the initial protostar mass and the strength and number of 
accretion outbursts
the observed lithium abundance of \futau seems in agreement with the predicted abundance
from these models. 
However, in strong contrast to the observation for \futau, the inclusion of accretion
in evolutionary models produces an \underline{under}luminosity in the HR diagram 
\citep{Baraffe09.0}. Special cases of these models evolve through a high-luminosity phase
but this phase is very short-lived and the probability of observing its realization in
nature is rather low. 
Moreover, 
\futau is the only known VLM YSO with strong lithium under-abundance that is a confirmed
accretor. 

A strong underabundance of lithium with respect to the cosmic abundance has also
been found by \cite{Johnas07.0} who modeled the high-resolution spectra of several young
VLM objects in the Cha\,I star forming region with synthetic atmospheres. Their targets
have spectral types M6/M7 and lithium abundances A(Li) $< 2.0$, just as \futau. This 
seems to imply that our understanding of lithium depletion or line formation 
in VLM objects is incomplete.

\section{Summary and conclusions}\label{sect:conclusions}

We have analysed in detail an X-Shooter spectrum of \futau which 
has allowed us to redetermine its stellar parameters and has provided a wealth of new 
information on activity, accretion and outflow diagnostics.
This allows us to exclude accretion luminosity and extinction 
as cause of the overluminosity of \futau in the HR diagram. 
An inflated radius as a result of magnetic activity 
leading to an underestimate of $T_{\rm eff}$ and $M_*$ might be responsible for it. 
Strong activity is consistent with the high X-ray luminosity of \futau.
A combination of magnetic activity and very young age for the FU\,Tau binary may be 
responsible for its HR diagram location. 
Another possibility is a closer distance but this is hard to explain due to the association
of \futau with the Barnard\,215 dark cloud that is believed to be part of the Taurus complex. 
The possibility of \futau being an unresolved binary must be taken into account. This
would bring \futau in better agreement with other luminous Taurus members and possibly 
with the $RV$ distribution of Taurus.  
High-resolution spectroscopy is needed
to search for a putative companion and to verify the unexpected evidence for lithium depletion. 
However, even with the available information the startling properties of \futau makes it 
a benchmark object to push on the further development of PMS evolutionary models.

\begin{acknowledgements}
JMA, BS and EC wish to thank G.Attusino for his help with the rapid completion of the
manuscript. We appreciated the interesting discussions with M.Bate. 
We thank the ESO staff for their support during the observations. We also appreciate the  
support of P. Goldoni, A. Modigliani and G. Cupani in the use of the X-Shooter pipeline.
AS wishes to acknowledge 
funding by the Science Foundation Ireland through grant no. 10/RFP/AST2780. 
\end{acknowledgements}

\bibliographystyle{aa} 
\bibliography{futauXS_v2}

\end{document}